\RequirePackage{silence}
\documentclass[12pt]{article}

\usepackage[top=2.0cm,bottom=2.3cm,left=2.54cm,right=2.54cm]{geometry}
\usepackage{mathrsfs,amsmath,amsfonts,mathtools,amssymb,wasysym,graphicx,subfigure,caption,rotating,comment,booktabs,cite}
\usepackage{bm}
\usepackage{color}  
\usepackage{extarrows}
\usepackage{appendix}

\usepackage[hyperindex=true,
        pdfstartview=FitH,
        bookmarksnumbered=true,
        bookmarksopen=true,
        citecolor=blue,
        linkcolor=blue,
        colorlinks=true,
        pdfborder=001,
        unicode]{hyperref}

\allowdisplaybreaks
\DeclareMathOperator{\e}{e}
\newcommand{\rd}{{\rm d}}
\newcommand{\ie}{\emph{i.e. }}
\newcommand{\eg}{\emph{e.g. }}

\renewcommand\({\left(}
\renewcommand\){\right)}

\newcommand{\blue}[1]{\textcolor{blue}{#1}}

\renewcommand{\blue}[1]{\textcolor{black}{#1}}

\newcommand{\ind}[2]{^{#1}{}_{#2}}

\newcommand{\add}{\text{add}}
\newcommand{\fa}{\mathfrak{a}}
\newcommand{\fb}{\mathfrak{b}}

\newcommand{\re}{{\rm e}}
\newcommand{\pfrac}[2]{\frac{\partial #1}{\partial #2}}

\newtheorem{theorem}{Theorem}[section]

\begin{document}

\title{Fluctuation theorems in general relativistic \\stochastic thermodynamics}

\author{
Yifan Cai\thanks{{\em email}: \href{mailto:caiyifan@mail.nankai.edu.cn}
{caiyifan@mail.nankai.edu.cn}},~
Tao Wang\thanks{
{\em email}: \href{mailto:taowang@mail.nankai.edu.cn}{taowang@mail.nankai.edu.cn}},~
and Liu Zhao\thanks{Corresponding author, {\em email}: 
\href{mailto:lzhao@nankai.edu.cn}{lzhao@nankai.edu.cn}}\\
School of Physics, Nankai University, Tianjin 300071, China}

\date{}
\maketitle

\begin{abstract}
Based on the recently proposed framework of general relativistic stochastic
mechanics [{\em J. Stat. Phys.}, 190:193, 2023; {\em J. Stat. Phys.}, 190:181, 2023]
and stochastic thermodynamics [{\em SciPost Physics Core} 7, 082, 2024]
at the ensemble level, this work focuses on general relativistic stochastic thermodynamics 
at the trajectory level. The first law of stochastic thermodynamics 
is reformulated and the fluctuation theorems are proved on this level,  
with emphasis on maintaining fully general covariance and on the choice of observers.
\vspace{1em}

\end{abstract}

\section{Introduction}

One of the central problems in modern statistical physics is the origin of irreversibility 
in macroscopic and mesoscopic systems. This problem can be traced back to Boltzmann's 
efforts in proving the second law of thermodynamics starting from deterministic 
mechanics, which has resulted in the well-known H-theorem \cite{Boltzmann1970}. 
However, the debates about the validity of (the assumptions of) the H-theorem 
have lasted for more than a century. The most acute criticism of the H-theorem is 
reflected by the Loschmidt paradox \cite{ehrenfest1911conceptual,wu1975boltzmann}, 
which roughly states that the macroscopic arrow of time cannot possibly arise from the 
underlying microscopic mechanics obeying time reversal symmetry (TRS). This paradox remained
unresolved until the 1990s, when numerous works \cite{evans1994equilibrium, 
gallavotti1995dynamical, crooks1999entropy} emerged, revealing that the forward
and reversed processes are not probabilistically equally likely, provided a certain 
dissipative effect exists on the mechanical level. These results, known as 
fluctuation theorems, largely resolved the debates and paradoxes related to the H-theorem. 
In particular, the most questioned molecular chaos hypothesis adopted in proving 
the H-theorem is completely avoided in proving the fluctuation theorems. 
 
Almost at the same time, Sekimoto \cite{sekimoto1998langevin} established the first 
law of stochastic thermodynamics on the trajectory level by use of the Langevin equation. 
This formulation establishes a connection between stochastic mechanics and fluctuation
theorems. In 2005, Seifert \cite{seifert2005entropy} presented a version of 
fluctuation theorem based on the overdamped Langevin equation. Subsequently, 
several fluctuation theorems based on non-relativistic stochastic mechanics were presented 
\cite{imparato2006fluctuation,chernyak2006path,ohkuma2007fluctuation,cai2024fluctuation},  
making it clear that stochastic mechanics provides an ideal starting point 
for constructing fluctuation theorems and interpreting the origin of irreversibility.

Most of the works mentioned above were carried out in the non-relativistic regime. 
Nowadays, it is widely acknowledged that thermodynamics and relativity are 
both concentrated on the universal principles that every physical 
system must obey. It is important to establish fluctuation theorems based on 
these universal principle theories. However, since the spacetime 
symmetry in relativity imposes stronger protection of time-reversal invariance, 
the extension of fluctuation theorems to the relativistic regime proves to be 
more difficult. The central difficulty lies in 
how to incorporate the breaking of time reversal invariance while still maintaining 
relativistic covariance. Refs. \cite{FINGERLE2007696,
cleuren2008fluctuation,fei2019quantum,teixido2020first,pal2020stochastic,pei2024special} 
considered the extension of fluctuation theorems to the special relativistic regime. 
However, the long awaited general relativistic extension is still beyond our 
ability to understand.

Recently, we developed a framework for dealing with stochastic mechanics on 
curved spacetime, and we investigated the general relativistic stochastic thermodynamics 
based on this framework \cite{cai2023relativistic,cai2023relativistic2,wang2023general}. 
Meanwhile, we also established a version of a fluctuation theorem on an arbitrary 
curved Riemannian manifold \cite{cai2024fluctuation}. The aim of the present work is to
employ the framework established in \cite{cai2023relativistic,
cai2023relativistic2,wang2023general} and make use of the technique introduced in 
\cite{cai2024fluctuation} to construct a version of a fluctuation theorem based 
on a fully general relativistic description of stochastic mechanics and 
stochastic thermodynamics. 

In our framework of relativistic stochastic mechanics \cite{cai2023relativistic,
cai2023relativistic2}, it is important to liberate the observer from the coordinate 
system. Rather than fixing the zeroth component of the coordinate system, we utilize 
the observer's proper time $t$ to label the configuration space $\mathcal S_t$ 
and the space of microstates $\Sigma_t$, enabling our theory to possess 
general covariance. For this purpose, we first clarify the geometry of the space of 
$\mathcal S_t$ and $\Sigma_t$ in Sec.~\ref{Sec:review}. To make the construction 
more self-contained, we provide a brief review of the basics of our framework 
of relativistic stochastic mechanics and relativistic stochastic thermodynamics in 
Sec.~\ref{Sec:Langevin}, and the first law of relativistic 
stochastic thermodynamics at the trajectory level is established in this section also. 
The separation of the observer from the coordinate system is also important 
in describing the time reversal transformation (TRT) in curved spacetime. 
In contrast to the usual practice in special relativistic theories (including 
special relativistic field theories) in which the TRT is often described as a 
coordinate transformation $(t,x^i)\mapsto (-t,x^i)$, our general covariant framework 
calls for interpreting the TRT as a transformation from a future-directed observer 
to a past-directed observer. This concept is elucidated in detail in 
Sec.~\ref{Sec:time-reversal}. Reference \cite{cai2023relativistic} presents a method 
to establish the covariant relativistic Langevin equation from the perspective 
of the observer. Based on this approach, the forward and reversed processes are defined, 
respectively, as stochastic processes from the perspectives of the 
future-directed and past-directed observers. In Sec.~\ref{Sec:fluctuation}, 
we demonstrate that the forward and reversed processes possess distinct 
probabilities, thus breaking the TRS and leading to 
a version of a fluctuation theorem on a curved spacetime manifold. 
Finally, in Sec.~\ref{Sec:conclusion}, we provide brief concluding remarks. 

We maintain the notations and conventions consistent with Refs.
\cite{cai2023relativistic,cai2023relativistic2}. To distinguish random 
variables from their realizations, the former are labeled with extra tildes. 
For instance, $\tilde x$ represents a random variable, while $x$ denotes its realization. 
Several manifolds of different dimensions will be relevant in our discussion. These 
include the $(d+1)$-dimensional spacetime manifold $\mathcal M$
with metric $g_{\mu\nu}(x)$ of signature $(-,+,\cdots,+)$, 
its tangent bundle $T\mathcal M$ of dimension $(2d+2)$, and certain submanifolds 
within the tangent bundle. To distinguish tensors on these different manifolds, 
we introduce distinct indices. Lower-case Greek letters, such as $\alpha, \beta, 
\mu, \nu, \rho, \ldots$, are used as concrete indices and range from $0$ to $d$.
Lower-case Latin letters, such as $i, j, k, l, m, \ldots$ are also used as concrete indices, 
which range from $1$ to $d$. Lastly, lower-case Latin letters $a, b, c,  \ldots$ 
are used as abstract indices. This paper is intended to be as mathematically rigorous 
as possible. A more concise summary of the main results can be found in 
Ref.\cite{CWZ-short}.

\section{Geometry of the space of microstates}
\label{Sec:review}

Statistical physics is built on top of the space of microstates. 
For systems consisting of classical massive particles, 
the space of microstates can be subdivided into configuration space 
and momentum space. In a relativistic context, configuration space 
is a subspace of the spacetime manifold 
$\mathcal M$ consisting of simultaneous events at a given instance of time, 
while the momentum space for each individual particle is a subspace of the 
tangent (or cotangent) space of the 
spacetime at a given event which obeys the mass shell condition
\begin{align}
\mathcal H(x,p):=g_{\mu\nu}(x)p^\mu p^\nu+m^2=0.
\end{align}
It is important to remember that, 
due to the non-degeneracy of the spacetime metric $g_{\mu\nu}(x)$, 
the tangent and cotangent spaces are dual to each other, and both can  
describe the momentum space equally well. In this work, we adopt the tangent space description. 
When cotangent space variables appear, they are considered to be 
linear functions of the tangent space variables, \eg $p_\mu = g_{\mu\nu}(x)p^\nu$.

In principle, the space of microstates for a single particle 
should be considered as a submanifold of 
the mass shell bundle
\begin{align}
\Gamma_m:=\{(x,p)\in T\mathcal M| \mathcal{H}(x,p)=0 \},
\end{align}
with the configuration space taken to be a subspace of the spacetime 
manifold $\mathcal M$ consisting of {\em simultaneous events}. To clarify the 
concept of simultaneous events, we need to introduce an arbitrary observer field 
that is encoded by a normalized timelike vector field $Z^\mu$ 
obeying $g_{\mu\nu}Z^\mu Z^\nu
=-1$. For convenience, we shall refer to this arbitrary observer field as {\em Alice}. 
When considering the motion of a Brownian particle inside a heat reservoir, 
there is a particular observer field that comoves with the reservoir. 
This particular observer field will be referred to as {\em Bob}.

Consider the worldline $x_\tau$ of a massive relativistic particle of which $\tau$ is  
its proper time. If the time orientations of Alice and the particle align, 
\ie $g_{\mu\nu} p^\mu_\tau Z^\nu <0$, where $p^\mu_\tau:=m\rd x^\mu_\tau/\rd \tau$, 
the part of the mass shell bundle in which the phase trajectory lies is defined as 
the {\em future mass shell bundle relative to Alice} and is denoted as 
$\Gamma^+_m$,
\begin{align}
\Gamma^+_m:=\{(x,p)\in T\mathcal M| \mathcal{H}(x,p)=0,\ g_{\mu\nu} p^\mu Z^\nu <0\}.
\end{align}
For notational convenience, the phase trajectory is denoted as $X_\tau=(x_\tau,p_\tau)$, 
which is the uplift of the particle's worldline $x_\tau$ into the 
bundle $\Gamma^+_m$.

\begin{figure}[h]
    \centering
    \includegraphics[scale=1.0]{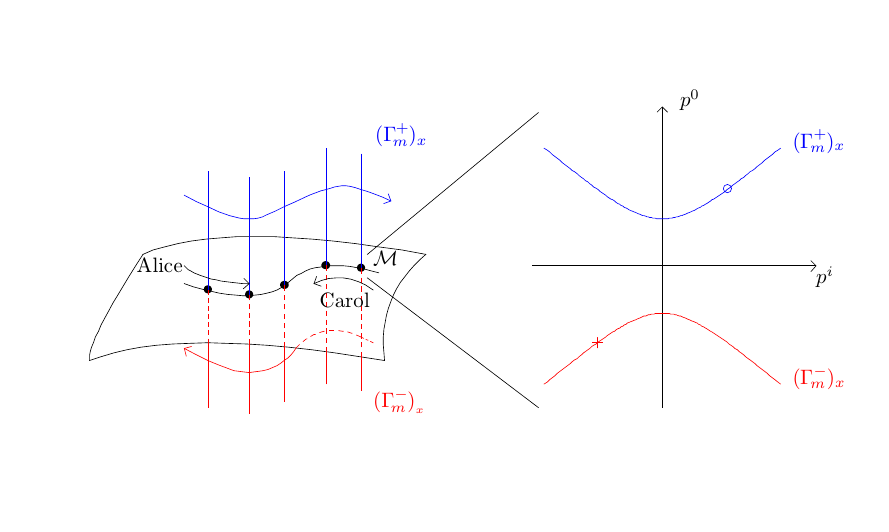}
    \caption{The worldline of a particle is lifted to different regions of 
    the mass shell bundle by different observers. }
    \label{fig1}
\end{figure}

If we consider another observer field with proper velocity $C^\mu=-Z^\mu$, referred 
to as {\em Carol}, whose time orientation is opposite to that of Alice, 
the definition of the future mass shell bundle 
relative to Carol will be opposite to that relative to Alice (see Fig.\ref{fig1}). 
To avoid confusion, 
we will designate Alice as the future-directed observer and Carol as the 
past-directed observer, and throughout this paper, 
the future and past mass shell bundles are always defined relative to Alice. 
This designation is arbitrary because, in the presence of TRS, 
the future- and the past-directed observer fields are indistinguishable 
on the level of deterministic mechanics.

The future mass shell bundle is also a fiber bundle based on $\mathcal M$. 
Its fiber, denoted by $(\Gamma^+_m)_x$, is the momentum space of the
relativistic particle. We use calligraphy letters, such as $\mathcal F$, $\mathcal R$ 
and $\mathcal K$, to denote tensors on $(\Gamma_m^+)_x$, and the cursive letters, 
such as $\mathscr N$, $\mathscr Z$ and $\mathscr L$, to denote 
tensors on $\Gamma_m^+$.

The future mass shell bundle is still not the space of microstates, 
because the base manifold $\mathcal M$ is not the configuration space. 
There are different ways to drop the temporal dimension of $\mathcal M$, e.g. 
either by fixing the zeroth component $x^0$ of coordinates or by 
fixing the proper time $t$ of Alice. The first approach lacks apparent general 
covariance. The core idea of our framework for relativistic stochastic 
mechanics is to keep the manifest general 
covariance and emphasize the role of observer choice. The details can be found 
in Ref.\cite{cai2023relativistic}. Here we list some of the key 
points:
\begin{enumerate}
\item The proper time $t$ of Alice can be extended into a scalar field $t(x)$ 
on $\mathcal{M}$;
\item The configuration space relative to Alice is defined as a 
constant proper time slice of $\mathcal M$, \ie a spacelike hypersurface  
$\mathcal{S}_t:=\{x\in \mathcal{M}|t(x)=t\}$ in $\mathcal M$. Moreover, 
the proper velocity field $Z^\mu$ of Alice is the unit normal vector 
field of $\mathcal{S}_t$;
\item The space of microstates is defined as a hypersurface in $\Gamma^+_m$ 
as a constant time slice, \ie $\Sigma_t^+:=\{(x,p)\in\Gamma^+_m|t(x)=t\}$;
\item Since the gradient of $t$ must be a normal co-vector of $\mathcal{S}_t$, 
$Z_\mu$ must be collinear with $\nabla_\mu t$. 
Denoting $|\nabla_\mu t|$ as $\lambda$, we have $\partial_\mu t=-\lambda Z_\mu$.
\end{enumerate}
Since $Z$ is the unit future-directed normal vector field of $\mathcal S_t$, 
the volume element on $\mathcal S_t$ can be written as the interior derivative 
of the volume element on $\mathcal{M}$ along $Z^\mu$:
\begin{align}
    \eta_{\mathcal{S}_t}:=\iota_{Z}\eta_{\mathcal M},
\end{align}
where $\iota$ represents interior derivative, and $\eta_{\mathcal M}
=g^{1/2}\rd x^0\wedge\rd x^1\wedge\cdots\wedge\rd x^d$ is the volume element on $\mathcal M$, 
with $g=|{\rm det}(g_{\mu\nu})|$.  
Since $\iota_{Z}(\rd x^\mu)=Z^\mu$ and $\rd x^0=-\partial_i t\rd x^i/\partial_0 t$ 
on $\mathcal S_t$, we have
\begin{align}\label{volume-configuration}
    \eta_{\mathcal S_t}&=g^{1/2}\iota_Z(\rd x^0\wedge\rd x^1\wedge\cdots\wedge\rd x^d)\notag\\
    &=g^{1/2}\sum_{\rho=0}^d (-1)^{\rho}\iota_{Z}(\rd x^\rho)
    \rd x^{0}\wedge \cdots \wedge\rd x^{\rho-1}\wedge\rd x^{\rho+1}\wedge\cdots\wedge\rd x^{d}\notag\\
    &=g^{1/2}Z^0 \rd x^1\wedge\cdots\wedge\rd x^d-g^{1/2}\sum_{i=1}^{d}Z^i \rd x^1\wedge\cdots\wedge\rd x^0\wedge\cdots\wedge\rd x^d\notag\\
    &=g^{1/2}\left[Z^0+\frac{1}{\partial_0 t}\partial_i t Z^i \right]\rd x^1\wedge\cdots\wedge\rd x^d\notag\\
    &=-\frac{\lambda g^{1/2}}{\partial_0t} \rd x^1\wedge\cdots\wedge\rd x^d.
\end{align}

We shall also need to make use of the volume elements on $\Gamma^+_m$ and $\Sigma_t^+$. 
Since both of them are submanifolds of $T\mathcal M$, it is appropriate to begin 
from the geometry of $T\mathcal M$ \cite{sarbach2014geo}. The non-degenerate 
metric on $T\mathcal M$ is known as the Sasaki metric 
\cite{sasaki1958differential} $\hat{g}_{ab}$, which is determined by the 
metric $g_{\mu\nu}$ of the base manifold $\mathcal{M}$,
\begin{align}\label{sasaki-metric}
\hat{g}_{ab}:=g_{\mu\nu}\rd x\ind{\mu}{a} \rd x\ind{\nu}{b}
+g_{\mu\nu}\theta\ind{\mu}{a}\theta\ind{\nu}{b},\quad
\hat{g}^{ab}:=g^{\mu\nu}e_{\mu}{}^a e_{\nu}{}^b
+g^{\mu\nu}\(\pfrac{}{p^\mu}\)^a\(\pfrac{}{p^\nu}\)^b,
\end{align}
where 
\[
\theta^{\mu}:=\rd p^{\mu}+\varGamma^{\mu}{}_{\alpha\beta}p^{\alpha} \rd x^{\beta}, 
\quad
e_{\mu}:=\pfrac{}{x^{\mu}}-\varGamma^{\alpha}{}_{\mu\beta}
p^{\beta}\pfrac{}{p^\alpha},
\]
and $\varGamma^{\mu}{}_{\alpha\beta}$ is
the Christoffel connection associated with $g_{\mu\nu}$. 
The corresponding volume element reads
\begin{align}\label{volumeelement}
\eta_{T\mathcal M}&=g~\rd x^0\wedge\rd x^1\wedge\cdots\wedge\rd x^d\wedge \theta^0\wedge\cdots
\wedge\theta^d\notag\\
&=g~\rd x^0\wedge\rd x^1\wedge\cdots\wedge\rd x^d\wedge\rd p^0\wedge\cdots\wedge\rd p^d.
\end{align}

As a hypersurface in $T\mathcal M$, $\Gamma^+_m$ has 
the unit normal (co)vector 
\begin{align}
\hat N_{a}=\frac{1}{|\rd \mathcal H|}\rd \mathcal H_{a}
=\frac{p_\mu}{m}\theta\ind{\mu}{a},
\qquad 
\hat N^a=\hat g^{ab}\hat N_{b}=\frac{p^\mu}{m}\(\pfrac{}{p^\mu}\)^a,
\end{align} 
giving rise to the induced metric
\begin{align}\label{metric-mass-shell-bundle}
\hat h_{ab}&:=\hat g_{ab}+\hat N_{a} \hat N_{b}
=g_{\mu\nu}\rd x\ind{\mu}{a}\rd x\ind{\nu}{b}
+\Delta_{\mu\nu}(p)\theta\ind{\mu}{a}\theta\ind{\nu}{b},\\
\Delta_{\mu\nu}(p) &= g_{\mu\nu}+\frac{1}{m^2} p_\mu p_\nu,
\end{align}
where the second term is the induced metric on $(\Gamma^+_m)_x$:
\begin{align}
h_{ab}:=\Delta_{\mu\nu}(p)\theta\ind{\mu}{a}\theta\ind{\nu}{b}.
\label{hab}
\end{align}
Consequently, we get the volume element 
\begin{align}
\eta_{\Gamma^+_m}&:=\iota_{\hat N}\eta_{T\mathcal M}
=-\frac{m}{p_0}g~\rd x^0\wedge\cdots\wedge\rd x^d\wedge\rd p^1\wedge\cdots\wedge\rd p^d.
\end{align}
It is easy to see that the volume element $\eta_{\Gamma^+_m}$ can be factorized 
into the wedge product of $\eta_{\mathcal M}$ and $\eta_{(\Gamma^+_m)_x}$,
\begin{align}
\eta_{\Gamma^+_m}=\eta_{\mathcal M}\wedge\eta_{(\Gamma_m^+)_x},
\end{align}
where 
\begin{align}
\eta_{(\Gamma^+_m)_x}:=-\frac{m}{p_0}g^{1/2}\rd p^1\wedge\cdots\wedge \rd p^d
\end{align}
is the volume element on the fiber space $(\Gamma^+_m)_x$. 
Finally, since $\Sigma^+_t=\bigcup_{x\in\mathcal{S}_t}(\Gamma_m^+)_x$, the volume element 
on $\Sigma^+_t$ also has a factorized form,
\begin{align}\label{volume_state_space}
\eta_{\Sigma^+_t}=\eta_{\mathcal{S}_t}\wedge\eta_{(\Gamma_m^+)_x}
=\frac{m\lambda g}{p_0\partial_0 t}\rd x^1\wedge\cdots\rd x^d
\wedge\rd p^1\wedge\cdots\wedge\rd p^d.
\end{align}

\section{General relativistic Langevin systems}
\label{Sec:Langevin}

This section is intended for a brief review of the framework for general 
relativistic stochastic mechanics and thermodynamics \cite{cai2023relativistic, cai2023relativistic2, wang2023general} in order to fix the notations and 
make the forthcoming presentation for the
proof of fluctuation theorem self-contained.

\subsection{Covariant Langevin equations}

Let us consider a relativistic Brownian particle carrying an electric charge $q$ and moving 
in a heat reservoir residing in the curved spacetime $\mathcal M$ and subjected to 
an external electromagnetic field $F=F_{\mu\nu} \rd x^\mu\wedge \rd x^\nu$. 
\blue{We assume that the heat reservoir has already reached thermal equilibrium, 
hence there is no difference between the Eckart frame and the Landau frame in 
defining the proper velocity $U^\mu$ of the reservoir.}
A version of the corresponding general covariant Langevin equation 
(referred to as LE$_\tau$) that takes the proper time $\tau$ of the particle 
as an evolution parameter reads
\begin{align}\label{LEtau-em-1}
\rd \tilde{x}_{\tau}^{\mu}&=\frac{\tilde{p}_{\tau}^{\mu}}{m}\rd \tau,\\
\label{LEtau-em-2}
\rd \tilde{p}_{\tau}^{\mu}&=\xi^\mu_\tau \rd\tau
+\mathcal{F}^\mu_{\text{dp}} \rd \tau
+\mathcal{F}^\mu_{\text{em}} \rd \tau
-\frac{1}{m}\varGamma\ind{\mu}{\alpha\beta}\tilde{p}_{\tau}^{\alpha}
\tilde{p}_{\tau}^{\beta}\rd \tau,
\end{align}
where $\mathcal{F}^\mu_{\text{dp}} := \mathcal{K}^{\mu\nu}U_{\nu}$ 
is the damping force with $\mathcal K^{\mu\nu}$ being the damping coefficient 
which transforms as a tensor under general coordinate transformations, 
$\mathcal{F}^\mu_{\text{em}} := \frac{q}{m}F^{\mu}{}_{\nu}\tilde{p}^{\nu}_{\tau}$
is the electromagnetic force, and 
\begin{align}
\xi_\tau^\mu:=\mathcal R\ind{\mu}{\fa}\circ_S\rd \tilde w^\fa_\tau/\rd \tau
+\mathcal{F}_\add^\mu
\end{align}
is the stochastic force, which consists of a random force term 
$\mathcal R\ind{\mu}{\fa}\circ_S\rd \tilde w^\fa_\tau/\rd \tau$ encoding the 
Stratonovich type coupling between the stochastic amplitudes $\mathcal R\ind{\mu}{\fa}$ 
with a set of independent Gaussian noises $\rd \tilde w^\fa_\tau$ 
obeying the probability distribution 
\begin{align}
\Pr[\rd \tilde w^\fa_\tau=\rd w^\fa]
=\frac{1}{(2\pi\rd \tau)^{d/2}}\exp\left(-\frac{1}{2}\frac{\delta_{\fa\fb}\rd w^\fa\rd w^\fb}
{\rd \tau}\right),
\end{align}
and an additional stochastic force term 
\begin{align}
\mathcal F_\add^\mu=\frac{\delta^{\fa\fb}}{2}
\mathcal R\ind{\mu}{\fa}\nabla^{(h)}_i \mathcal R\ind{i}{\fb}
\label{Fadd}
\end{align}
in which $\nabla^{(h)}_i$ denotes the spatial components of the covariant derivative 
associated with the metric \eqref{hab} on the momentum space. 
Both the random force and the additional stochastic force terms arise 
from the interaction of the Brownian particle with the heat reservoir. 
The additional stochastic force is required in order for the Brownian 
particle to be able to reach equilibrium 
in the long time limit \cite{klimontovich1994nonlinear}. 
Each component of $\mathcal R\ind{\mu}{\fa}$ is assumed to be smoothly dependent 
on the coordinates on $\Sigma_t^+$, and 
for each fixed $\fa=1,2,\cdots d$, $\mathcal R\ind{\mu}{\fa}$ transforms as a
vector under general coordinate transformations. 
As was done in Refs.~\cite{cai2023relativistic,cai2023relativistic2}, 
we use tilded and un-tilded symbols to denote the random variables 
and their realizations.

Although LE$_\tau$ is perfectly generally covariant and encodes all necessary 
factors that affect the motion of the Brownian particle, there are still some 
drawbacks that call for an alternative version of the covariant Langevin equation. 
The problem is connected to the choice of evolution parameter $\tau$. 
Since
\begin{align}
\rd t=\partial_\mu t \rd \tilde x^\mu
=-\lambda Z_\mu\rd \tilde x^\mu=-\lambda Z_\mu \frac{\rd \tilde x^\mu}{\rd \tau}\rd \tau
=-\lambda \frac{Z_\mu \tilde p^\mu}{m}\rd \tau
= \gamma(\tilde x,\tilde p)\rd\tau,
\label{dtvsdtau}
\end{align}
we have $\rd \tau =\gamma^{-1}(\tilde x,\tilde p)\rd t$. Therefore, 
from the perspective of the observer Alice, the proper time
$\tau$ of the Brownian particle becomes a random variable. To avoid this inconvenience, 
a reparametrization scheme is adopted in \cite{cai2023relativistic}, with
\[
\tilde X_\tau=(\tilde x_\tau, \tilde p_\tau) ~\mapsto~
\tilde Y_t =(\tilde y_t, \tilde k_t),
\]
where 
\[
\tilde y_t:=\tilde x_{\tilde \tau_t}, \quad 
\tilde k_t:=\tilde p_{\tilde \tau_t}.
\] 
This leads to the following alternative version of covariant Langevin equation 
which is referred to as LE$_t$ for short,
\begin{align}
\rd \tilde y_t^\mu&=\frac{\tilde k^\mu_t}{m}\gamma^{-1}\rd t,
\label{38}\\
\rd \tilde k_t^\mu&=\hat \xi_t^\mu \gamma^{-1}\rd t
+\mathcal{F}^\mu_{\text{dp}}\gamma^{-1}\rd t
+\mathcal{F}^\mu_{\text{em}}\gamma^{-1}\rd t
-\frac{1}{m}\varGamma\ind{\mu}{\alpha\beta}
\tilde k^\alpha_t\tilde k^\beta_t \gamma^{-1}\rd t.
\label{39}
\end{align}
The new stochastic force $\hat\xi^\mu_t$ reads
\begin{align}
\hat\xi^\mu_t:=\gamma^{1/2}\mathcal{R}\ind{\mu}{\fa}\circ_S\rd \tilde W_t^\fa/\rd t
+\mathcal{F}^\mu_\add-\frac{1}{2}\mathcal{D}^{\mu i}\gamma^{1/2}\nabla^{(h)}_i\gamma^{-1/2},
\end{align}
in which
\begin{align*}
\rd \tilde W^\mathfrak{a}_t
= \gamma^{1/2}(\tilde Y_t ) \rd \tilde w^\mathfrak{a}_{\tilde \tau_{t}} 
\end{align*}
are still Gaussian noises but with the variance changed from $\rd \tau$ to $\rd t$, 
and $\mathcal{D}^{\mu\nu}:= \mathcal R\ind{\mu}{\fa}
\delta^{\fa\fb}\mathcal R\ind{\nu}{\fb}$ is the diffusion tensor. 

In this work, we assume 
that the diffusion tensor has rank $d$, so that $\mathcal{D}^{ij}$ is a full-rank symmetric 
matrix. This also requires that $\mathcal R\ind{i}{\fa}$ is a $d\times d$ full-rank matrix. 
The above assumption is necessary and sufficient to ensure 
that the Brownian particle couples to the heat reservoir in every spatial direction.
Please keep in mind that the choice for the stochastic amplitudes $\mathcal R\ind{\mu}{\fa}$
is non-unique. Different choices correspond to different Langevin systems. 
The result of the present work does not require the explicit values for the 
stochastic amplitudes and should be valid for any choices obeying the above assumption.

\subsection{Reduced Fokker-Planck equation}

Using the diffusion operator method \cite{bakry2014analysis}, 
the reduced Fokker-Planck equation (RFPE) associated with LE$_\tau$ or LE$_t$ 
is obtained in Ref.~\cite{cai2023relativistic2},
\begin{align}\label{reduced-fp-em}
    \dfrac{1}{m}\mathscr{L}_{F}(\varphi)=\nabla_{i}^{(h)}\mathcal{I}^{i}[\varphi],
\end{align}
where 
\begin{align}\label{Liouville-I}
\mathscr{L}_F:= p^\mu e_\mu+q F^{\mu}{}_{\nu}p^\nu \pfrac{}{p^\mu}
\end{align}
is the Liouville vector field for a charged particle,
\begin{align} \label{Liouville-II}
\mathcal I[\varphi]=\left[ \frac{1}{2}\mathcal{D}^{ij} \nabla^{(h)}_j\varphi
-\mathcal{K}^{i\nu}U_\nu\varphi  \right] \frac{\partial }{\partial \breve p^i}
\end{align}
is a vector field which is connected to the heat transfer rate from the heat 
reservoir to the Brownian particle via \cite{cai2023relativistic,wang2023general}
\begin{align}
Q[\varphi]:=\int_{(\Gamma^+_m)_x}\eta_{(\Gamma^+_m)_x} Z_\nu \mathcal I^\nu[\varphi],
\label{heat-current}
\end{align}
and the definition for the derivative operator
$\displaystyle\frac{\partial }{\partial \breve p^i}$ is provided in
Appendix \ref{app:tangent-vector}. 
The round and square brackets around $\varphi$ have different meanings:
$\mathscr{L}_{F}(\varphi)$ represents the action of the vector field $\mathscr{L}_{F}$ on 
the scalar $\varphi$, while $\mathcal{I}^{i}[\varphi]$ implies that the 
vector field $\mathcal{I}^{i}$ is dependent on $\varphi$. Such convention will be used 
throughout this paper.

It is important to point out that
the one particle distribution function (1PDF) $\varphi$ appearing in the RFPE 
is {\em not} a probability distribution in $\Sigma^+_m$. To see this, 
we recall that the probability current associated with the RFPE \eqref{reduced-fp-em} is 
\begin{align}\label{probability-current}
\mathscr{J}[\varphi]=\frac{\varphi}{m}\mathscr{L}_F-\mathcal{I}[\varphi].
\end{align}
Therefore, the probability distribution function on $\Sigma^+_t$ should be
\begin{align}\label{pro-microstate}
f:=-{\mathscr{Z}_{a}}\mathscr{J}^a[\varphi]
=-\frac{1}{m}p^\mu Z_\mu\varphi=\gamma\lambda^{-1}\varphi,
\end{align}
where $\mathscr{Z}=Z^{\mu}e_{\mu}$ is the unit normal vector of $\Sigma_t^+$. 

Since $\mathcal I[\varphi]$ is proportional to  
the heat transfer rate, the condition for the 
Brownian particle to reach detailed thermal equilibrium with the reservoir is 
$\mathcal I[\varphi_{\text{eq}}]=0$, which yields the equilibrium distribution
\begin{align}\label{equilibrium-state}
\varphi_{\text{eq}}=\e^{-\alpha+\beta_\mu p^\mu},\quad \beta_\mu:=\beta U_\mu,
\end{align}
provided that the covariant Einstein relation
\begin{align}
\mathcal{D}^{\mu\nu} = 2\beta^{-1}\mathcal{K}^{\mu\nu}
\end{align}
holds and that $\alpha$ and $\beta_\mu$ obey the following equations
\begin{align}
\nabla_{\mu}\alpha+q\beta^{\nu}F_{\mu\nu}=0, \qquad
\nabla_{(\mu}\beta_{\nu)}=0, \label{killing}
\end{align}
which are simple consequences of the Liouville equation 
$\mathscr{L}_F(\varphi_{\text{eq}})=0$. Eq.~\eqref{killing} implies that $\beta_\mu$ 
is a Killing vector field, while the relation $\beta_\mu=\beta U_\mu$ implies 
that it is timelike. Therefore, $\beta_\mu$ must be timelike Killing. This leads to the 
conclusion that the existence of the equilibrium distribution \eqref{equilibrium-state}
requires the spacetime to be at least stationary.

As we have argued in Ref.~\cite{cai2023relativistic2},
the equilibrium state is intrinsic to the system which 
is independent of the choice of observer. However, the parameters that characterize
the equilibrium state is indeed observer-dependent. It has been shown 
\cite{hao2022relativistic,cai2023relativistic2} that the parameters $\alpha$ and $\beta$ are 
related to the chemical potential and the temperature observed by Bob via 
\begin{align} \label{tempbob}
\beta=\frac{1}{T_{\rm B}},\qquad
\alpha =-\frac{\mu_{\rm B}}{T_{\rm B}}.
\end{align}
The equilibrium distribution \eqref{equilibrium-state} is recognized to be 
precisely the J\"utnner distribution which is also obeyed by 
particles of the heat reservoir.

\subsection{Thermodynamic relations}

The definition of the energy of a charged relativistic particle is non-unique. 
For instance, both the kinematic momentum $p^\mu$ and the canonical momentum 
$P^\mu:=p^\mu+qA^\mu$ can be used for defining the energy \cite{landau1975classical}:
\begin{align}
E:=-Z_{\mu}p^{\mu},\qquad   H:=-Z_{\mu}P^{\mu}.
\end{align}
As an analogy of the non-relativistic case, $E$ can be viewed as kinematic energy 
and $H$ can be viewed as the sum of kinematic energy and electromagnetic potential 
energy $-qZ_\mu A^\mu$. The non-uniqueness for the definition of energy also 
appeared in the non-relativistic stochastic thermodynamics \cite{cai2024fluctuation}. 
However, such non-uniqueness does not affect the description 
of heat in the first law of stochastic thermodynamics. We will show that 
the same situation also occurs in the relativistic case.

Since the microstate of the Brownian particle is described by a set of random variables, 
the energy of the Brownian particle also depends on the same set of random variables,
\begin{align}
\tilde E_\tau:=E(\tilde x_\tau,\tilde p_\tau),\qquad
\tilde H_\tau:=H(\tilde x_\tau,\tilde p_\tau).
\end{align}
Since LE$_\tau$ is a system of Stratonovich-type  
stochastic differential equations, the chain rule is available
\begin{align}
\rd \tilde E_\tau&=\pfrac{E}{x^\mu}\rd \tilde x^\mu_\tau
+\pfrac{E}{p^\mu}\rd \tilde p^\mu_\tau\notag\\
&=-Z_\mu\left[\xi_\tau^\mu+\mathcal F_{\text{dp}}^\mu \right]\rd \tau
-\frac{\tilde p_\tau^\mu \tilde p_\tau^\nu}{m}\nabla_\nu Z_\mu \rd \tau
-Z_\mu \mathcal F_{\text{em}}^\mu\rd \tau.
\end{align}
Similarly,
\begin{align}
\rd \tilde H_\tau&=\rd \tilde E_\tau
-\frac{q}{m}\pfrac{}{x^\nu}(Z_\mu A^\mu)\tilde p^\nu_\tau\rd \tau\notag\\
&=\rd \tilde E_\tau+Z_\mu\mathcal{F}^\mu_{\text{em}}\rd \tau
-\frac{q}{m}[A_\mu\nabla_\nu Z^\mu+Z^\mu\nabla_\mu A_\nu]\tilde p^\nu_\tau\rd \tau\notag\\
&=-Z_\mu\left[\xi_\tau^\mu+\mathcal F_{\text{dp}}^\mu \right]\rd \tau
-\frac{\tilde p_\tau^\mu \tilde p_\tau^\nu}{m}\nabla_\nu Z_\mu \rd \tau
-\frac{q}{m}\pounds_{Z}A_{\mu} \tilde p^\mu_\tau\rd \tau,
\end{align}
where $\pounds_{Z}A_{\mu}$ is the Lie derivative of $A^\mu$ along the vector field $Z^\mu$. 
If the electromagnetic field is controlled by an external protocol denoted by $\sigma$, 
the last term of the above equation can be rewritten as
\begin{align}
\rd_\sigma\tilde{\mathcal U}:=-\frac{q}{m}\pounds_{Z}A_{\mu} \tilde p^\mu_\tau\rd \tau.
\end{align}

In the realm of stochastic thermodynamics, the energy exchange between 
the Brownian particle and the heat reservoir is considered as heat, 
while the other part of the change of energy of the Brownian particle 
is considered as work. Therefore, the heat received by the Brownian particle 
from the heat reservoir is identified to be
\begin{align}
\rd \tilde{\mathcal Q}_\tau
:=-Z_\mu\left[\xi_\tau^\mu+\mathcal F_{\text{dp}}^\mu \right]\rd \tau,
\end{align}
and gravitational \cite{liu2021work} and electromagnetic works are identified respectively as
\begin{align}
&\rd \tilde{\mathcal P}_\tau
:=-\frac{\tilde p_\tau^\mu \tilde p_\tau^\nu}{m}\nabla_\nu Z_\mu \rd \tau,\\
&\rd \tilde{\mathcal W}_\tau:=-Z_\mu\mathcal{F}^\mu_{\text{em}}\rd \tau.
\end{align}
Therefore, the first law of relativistic stochastic thermodynamics 
can be realized either as
\begin{align}
\label{dE}
\rd \tilde E_\tau=\rd \tilde{\mathcal Q}_\tau+\rd \tilde{\mathcal P}_\tau
+\rd \tilde{\mathcal W}_\tau,
\end{align}
or as
\begin{align}
\label{dE1}
\rd \tilde H_\tau=\rd \tilde{\mathcal Q}_\tau+\rd \tilde{\mathcal P}_\tau
+\rd_\sigma\tilde{\mathcal U}.
\end{align}

The energy currents associated with the above two definitions of energy 
are presented as follows,
\begin{align}
E^{\mu}[\varphi]&:=\int_{{(\Gamma^+_m)_x}}\eta_{(\Gamma^+_m)_x}\frac{p^\mu}{m}\varphi E
=-Z_{\nu}T^{\mu\nu}[\varphi],
\\
H^{\mu}[\varphi]&:=\int_{{(\Gamma^+_m)_x}}\eta_{(\Gamma^+_m)_x}\frac{p^\mu}{m}\varphi H
=E^{\mu}[\varphi]-qZ_\nu A^{\nu}N^\mu[\varphi],
\end{align} 
wherein the energy-momentum tensor $T^{\mu\nu}[\varphi]$ reads
\begin{align}
T^{\mu\nu}[\varphi]
:=\int_{{(\Gamma^+_m)_x}}\eta_{(\Gamma^+_m)_x}\frac{p^\mu p^\nu}{m}\varphi.
\end{align}
In relativistic kinetic theory, the entropy current 
associated with classical non-degenerate particles 
is defined as \cite{cercignani2002relativistic}
\begin{align}\label{def-entropy-browian-particle}
S^\mu[\varphi]=-\int_{{(\Gamma^+_m)_x}}\eta_{(\Gamma^+_m)_x}
\frac{p^\mu}{m}\varphi\(\ln\varphi-1 \).
\end{align}
In the equilibrium distribution \eqref{equilibrium-state}, we have the following 
relation \cite{hao2022relativistic}:
\begin{align}\label{covariant-Euler}
S^{\mu}[\varphi_{\text{eq}}] 
&= \beta^\mu P  - T^{\mu \nu}[\varphi_{\text{eq}}]  \beta^\nu  
+ \alpha n^\mu[\varphi_{\text{eq}}],
\end{align}
where $\displaystyle
P:=\frac{1}{d}\Delta_{\mu\nu}(U) T^{\mu\nu}[\varphi_{\text{eq}}]$ is the pressure 
and $n^\mu[\varphi]$ is the particle current. Following a standard procedure as did 
in \cite{hao2022relativistic}, it can be shown that $ T^{\mu \nu}[\varphi_{\text{eq}}]$ and 
$n^\mu[\varphi_{\text{eq}}]$ can both be expressed in terms of the proper velocity $U^\mu$
of Bob: $n^\mu[\varphi_{\text{eq}}]$ is simply proportional to $U^\mu$, while 
$T^{\mu \nu}[\varphi_{\text{eq}}]$ takes the form of the energy-momentum tensor of 
a perfect fluid.

Now we would like to reinterpret eq.\eqref{covariant-Euler} in terms of the 
densities of various physical quantities as measured by Alice. This can be achieved by 
contracting each term in eq.\eqref{covariant-Euler} with $Z_\mu$. Recalling the 
fact that $\beta^\mu=\beta U^\mu$, the relationship between the proper velocities of 
Alice and Bob, can be written as
\begin{align}\label{decompose-Bob}
U^\mu=\gamma_{\rm A}(Z^\mu+z^\mu),\qquad \gamma_{\rm A}=-U^\mu Z_\mu,\qquad z^\mu Z_\mu=0,
\end{align}
where $\gamma_{\rm A}$ represents the 
local Lorentz factor arising from the relative motion between Alice and Bob. 
Consequently, we have
\begin{align}\label{Euler1}
e_{\rm A}&=(\gamma_{\rm A}\beta)^{-1} s_{\rm A}-\alpha(\gamma_{\rm A}\beta)^{-1}  n_{\rm A}- P+{\mathcal{T}}_{\rm A},
\end{align}
or alternatively
\begin{align}\label{Euler2}
h_{\rm A}&=(\gamma_{\rm A}\beta)^{-1} s_{\rm A}-[qZ_\mu A^\mu+\alpha(\gamma_{\rm A}\beta)^{-1} ] n_{\rm A}
- P+{\mathcal{T}}_{\rm A},
\end{align}
where 
\begin{align}
&e_{\rm A}:=T^{\mu\nu}[\varphi_{\text{eq}}]Z_\mu Z_\nu,
\quad
h_{\rm A}:= e_{\rm A}-q Z_\mu A^\mu n_{\rm A},
\quad
n_{\rm A}:=-Z_\mu n^\mu[\varphi_{\text{eq}}],\\
&s_{\rm A}:=-Z_\mu S^\mu[\varphi_{\text{eq}}],
\quad
{\mathcal{T}}_{\rm A}:=- T^{\rho\nu}[\varphi_{\text{eq}}]z_\mu Z_\nu.
\end{align}
$e_{\rm A}$ is the density of the energy $E$, $h_{\rm A}$ is 
the density of the energy $H$, $n_{\rm A}$ is the particle number density, 
$s_{\rm A}:=-Z_\mu S^\mu[\varphi_{\text{eq}}]$ is the entropy density, and 
${\mathcal{T}}_{\rm A}$ is the relative kinematic energy density arising 
from the relative motion between 
Alice and Bob. The subscript $\rm A$ in the notation for all these 
density quantities indicates that 
they are all defined with respect to the observer Alice. 
Equation \eqref{Euler1} can be viewed as the localized version of the Euler relation
in which the coefficients of $s_{\rm A}$ and $n_{\rm A}$ should be interpreted as 
the temperature and chemical potential as measured by Alice. Therefore we have 
\begin{align}\label{def-tem}
T_{\rm A}:=(\gamma_{\rm A}\beta)^{-1},\qquad \mu_{\rm A}:=-\alpha(\gamma_{\rm A}\beta)^{-1}.
\end{align}
Similarly, from eq.~\eqref{Euler2}, we can read off
\begin{align} \label{def-tem2}
\hat \mu_{\rm A}:= -[qZ_\mu A^\mu+\alpha(\gamma_{\rm AB}\beta)^{-1} ]
=\mu_{\rm A}-qZ_\mu A^\mu,
\end{align}
which is recognized to be the electrochemical potential. Inserting eq.\eqref{tempbob} into 
eq.~\eqref{def-tem}, the transformation rule for the temperature and chemical potential 
presented in Ref.\cite{hao2022relativistic} can be easily recovered,
\[
T_{\rm A}:=(\gamma_{\rm A})^{-1}T_{\rm B},\qquad
\mu_{\rm A}:=(\gamma_{\rm A})^{-1}\mu_{\rm B}.
\]
We can also substitute eqs.~\eqref{def-tem} and \eqref{def-tem2} 
into eqs.~\eqref{Euler1}-\eqref{Euler2} to make the appearance of the Euler relation 
simpler,
\begin{align}\label{Euler1-full}
e_{\rm A}&=T_{\rm A} s_{\rm A}+\mu_{\rm A} n_{\rm A}- P+{\mathcal{T}}_{\rm A},\\
\label{Euler2-full}
h_{\rm A}&=T_{\rm A} s_{\rm A}+\hat\mu_{\rm A}  n_{\rm A}- P+{\mathcal{T}}_{\rm A}.
\end{align}
The different choices of definition for the energy induce different chemical potentials, 
while such choices have no influence on the temperature. Putting these thermodynamic 
quantities into eq.~\eqref{equilibrium-state}, the equilibrium distribution 
can be rewritten as 
\begin{align}
\varphi_{\text{eq}}=\re^{(E-\mu_{\rm A}-z_\nu p^\nu)/T_{\rm A}}
=\re^{(H-\hat\mu_{\rm A}-z_\nu p^\nu)/T_{\rm A}}.
\end{align}

It remains to introduce the {\em trajectory entropy} for the Brownian particle, 
which plays an important role in the formulation of the fluctuation theorem. 
Denoting the phase trajectory of the Brownian particle as 
$\tilde Y_t=(\tilde y_t,\tilde k_t)$, the trajectory entropy
is defined as
\begin{align}\label{trajectory-entropy}
\tilde S_t=-\ln\varphi(\tilde Y_t),
\end{align}
The entropy production of the Brownian particle in the time interval 
$[t_I,t_F]$ is related to the trajectory entropy via
\begin{align}
\Delta S=\int_{\mathcal S_{t_F}}\eta_{\mathcal S_{t_F}} Z_\mu S^\mu[\varphi]
-\int_{\mathcal S_{t_I}}\eta_{\mathcal S_{t_I}} Z_\mu S^\mu[\varphi]
=\langle \tilde S_{t_F} \rangle-\langle \tilde S_{t_I} \rangle.
\end{align}
This part of the entropy production is also referred to as the trajectory 
entropy production.

\section{Time-reversal symmetry and its breaking}
\label{Sec:time-reversal}

Before delving into the construction of fluctuation theorems in the framework of 
general relativistic stochastic thermodynamics, it is crucial to accurately describe 
what the TRS is meant on curved spacetime. 

In ordinary textbooks on special relativistic field theories, 
the time-reversal transformation (TRT) is often represented by a coordinate 
transformation $(x^0,x^i)\mapsto (-x^0,x^i)$ with different transformation rules 
for temporal and spatial components of the coordinate bases
\begin{align}
\pfrac{}{x^0}\mapsto-\pfrac{}{x^0},\qquad \pfrac{}{x^i}\mapsto\pfrac{}{x^i}.
\end{align} 
Since $x^0$ serves as the evolution parameter, the spatial components $p^i$ of 
the momentum should reverse their signs, while the temporal component of the 
momentum remains unchanged, which corresponds to the particle's energy. 
Consequently, the transformation under TRT for the contraction of momentum 
components and the basis obeys
\begin{align}
p=p^0\pfrac{}{x^0}+p^i\pfrac{}{x^i}\mapsto -p^0\pfrac{}{x^0}-p^i\pfrac{}{x^i}=-p.
\end{align}

To maintain general covariance in our construction, it is better to {treat} the TRT 
as a change of the observer's time orientation, \ie from the perspective 
of Alice to that of Carol, rather than treating it as a coordinate transformation.
Meanwhile,  {we adopt the scalar expression like $-Z_\mu p^\mu$ for defining 
the energy of a particle instead of using the coordinate dependent definition $p^0$.}
Therefore, we employ an alternative representation of the TRT, 
which reverses the sign of full momentum (\ie $p^\mu\mapsto -p^\mu$) 
while keeping the spacetime coordinates intact. 
Such a representation implies the reversal of the particle's proper time derivative: 
$\rd/\rd \tau \mapsto -\rd/\rd \tau$. If Alice observes a particle evolving from 
$\tau_I$ to $\tau_F$ in its own proper time during the time interval $t_I \rightarrow t_F$, 
then Carol will observe the proper time of the charged particle evolving from $\tau_F$ 
to $\tau_I$.

\begin{figure}[h]
    \centering
    \includegraphics[scale=.7]{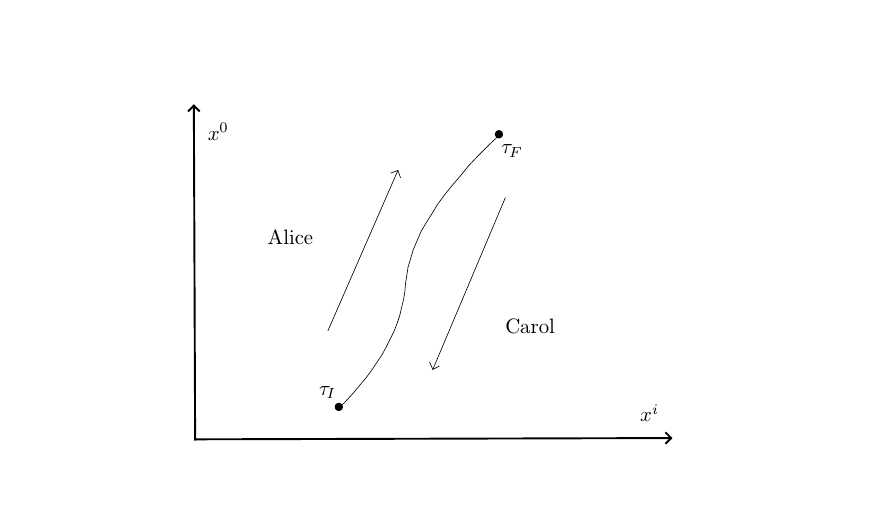}
    \caption{The evolution directions of the particle under the perspectives of Alice and Carol.}
    \label{fig2}
\end{figure}

Since the momentum is defined only in the local tangent space on curved spacetime, 
the TRT can be realized as an automorphism of $T_x\mathcal M$ (see Fig.\ref{fig2}):
\begin{align}
I_x: p\mapsto -p, \qquad\text{for }\forall p\in T_x\mathcal M.
\end{align} 
Furthermore, $I_x$ induces a homeomorphism 
between the future and past mass shell bundles,
\begin{align}
&I:\Gamma^+_m\to \Gamma^-_m,\\
&I:(x,p)\mapsto (x,I_x(p))=(x,-p).
\end{align}
Therefore, the phase trajectories lifted by Alice and Carol are on {two disconnected} 
regions of the mass shell bundle. For convenience, the trajectory lifted by Alice 
will be referred to as the \emph{forward trajectory}, {and the} one lifted by Carol 
will be referred to as the \emph{reversed trajectory}. 
For later reference, we also list the induced action (\ie pushforward) 
of the map $I$ on the coordinate basis for the vector field on $T\mathcal{M}$,
\begin{align}
I^*\pfrac{}{p^\mu}=-\pfrac{}{p^\mu},\qquad I^*\pfrac{}{x^\mu}=\pfrac{}{x^\mu}.
\end{align}

If we discretize the time interval $[t_I, t_F]$ into a sequence of $N+1$ 
equal-distance nodes $t_I=t_0<t_1<\cdots<t_N=t_F$, then the forward trajectory 
can be written as a sequence $Y_{[t]}:=(Y_0,Y_1,\cdots,Y_N)$ with $Y_n:=Y_{t_n}$. 
Consequently, the reversed trajectory can also be written as a sequence 
$Y_{[t]}^-:=(Y_0^-,Y_1^-,\cdots,Y_N^-)$ in which 
\begin{align}\label{relation-trajectory}
Y^-_n=I(Y_{N-n}).
\end{align}
The probability distribution $f=-p^\mu Z_\mu\varphi/m$ describes the distribution of 
the intersections of the stochastic trajectories with the space of microstates $\Sigma^+_t$, 
and it can be easily extended to the past mass shell bundle by use of the relationship 
between the forward and the reversed trajectories. Additionally, 
{as a metric-preserving transformation, the TRT $I$ preserves the volume 
elements on every submanifolds of the tangent bundle $T\mathcal M$ mentioned 
in Sec.~ \ref{Sec:review}. } Consequently, we have
\begin{align}\label{TRT-volume}
I^*\varphi=\varphi,\qquad I^*\eta_{\Gamma^+_m}=\eta_{\Gamma^-_m},
\qquad  I^*\eta_{(\Gamma^+_m)_x}=\eta_{(\Gamma^-_m)_x},
\qquad  I^*\eta_{\Sigma^+_t}=\eta_{\Sigma_t^-},
\end{align}
where $\Sigma_t^-$ is the space of microstates associated with the past directed 
observer, Carol. 

It remains to describe the TRT for the electromagnetic field. Consider the motion of a 
massive charged particle subjected to the electromagnetic field $F^{\mu\nu}$. 
The equation of motion reads
\begin{align}
\frac{\rd p^\mu}{\rd \tau} = \frac{q}{m} F^{\mu}{}_\nu p^\nu.
\end{align}
The TRS implies that the above equation of motion remains invariant. 
Since $p^\mu$ is odd and hence $\displaystyle\frac{\rd p^\mu}{\rd \tau}$ is even under the TRT,
the invariance of the equation of motion requires either
\begin{align}
q\mapsto q, \qquad F^{\mu\nu}\mapsto -F^{\mu\nu}
\end{align}
or 
\begin{align}
q\mapsto -q, \qquad F^{\mu\nu}\mapsto F^{\mu\nu}. 
\label{Feynman}
\end{align}
We will adopt the second convention which is known as Feynman's convention in the 
literature\cite{feynman1965nobel}. Therefore, we have
\begin{align}
I^*q = -q,\qquad I^* \mathcal{F}_{\text{em}}^\mu=\mathcal{F}_{\text{em}}^\mu.
\end{align}
Using the definition \eqref{Liouville-I}, it is not difficult to check that the 
induced action of the TRT on the Liouville vector field is given by 
\begin{align}
I^*\mathscr{L}_F=-\mathscr{L}_F.
\end{align}
Bringing all the above conventions together, we will see that the equations for 
the phase trajectory of a massive charged particle moving in a generic curved spacetime
and subjected to an external electromagnetic field remains invariant under the TRT,
\begin{align}\label{5.33}
\frac{\rd x^\mu}{\rd \tau}&=\frac{p^\mu}{m},\\
\label{5.34}
\frac{\rd p^\mu}{\rd \tau}&=\frac{q}{m}F^{\mu}{}_{\nu}p^\nu
-\frac{1}{m}\varGamma\ind{\mu}{\alpha\beta}p^\alpha p^\beta.
\end{align}

Let us now bring the above system into a broader picture. Consider 
a scenario in which there are a great number of massive charged particles of different 
masses and charges moving together. The electromagnetic field is produced by the 
particles themselves and the spacetime geometry is 
determined by the masses and charges of the particles. 
We assume that the system consists of different species of particles, 
and each species carries different mass $m_s$ and charge $q_s$
and obeys a different distribution $\Phi_s$, which are differentiated from each other by 
the suffix $s$. We also assume that the distributions $\Phi_s$
are invariant under the TRT, \ie
$I^*\Phi_s =\Phi_s$. Then the total electric current 
\begin{align}\label{electric-current}
J^\mu=\sum_s \int \eta_{(\Gamma_{m_s}^+)_x}\frac{q_s p^\mu_s}{m_s}\Phi_s
\end{align}
as well as the energy-momentum tensor 
contributed by the particles
\begin{align}\label{momentum-tensor-particles}
T^{\mu\nu}_{\text{pa}}=\sum_s \int \eta_{(\Gamma_{m_s}^+)_x}
\frac{ p^\mu_s p^\nu_s}{m_s}\Phi_s
\end{align}
should both be TRT invariants. Consequently, the Maxwell equation 
\begin{align}\label{maxwell}
\nabla_\nu F^{\mu\nu}=J^\mu
\end{align}
that determines the electromagnetic field as well as the Einstein equation
\begin{align}\label{5.37}
R^{\mu\nu}-\frac{1}{2}g^{\mu\nu} R=8\pi G\(T^{\mu\nu}_{\text{pa}}+T^{\mu\nu}_{\text{em}}\)
\end{align}
that determines the spacetime geometry should all be invariant under the TRT, 
wherein 
\begin{align}\label{momentum-tensor-electric}
T^{\mu\nu}_{\text{em}}= F^{\mu \rho} F\ind{\nu}{\rho}
-\frac{1}{4}g^{\mu\nu}F^{\rho\sigma}F_{\rho\sigma}
\end{align}
is the energy-momentum tensor of the electromagnetic field, whose TRT invariance 
is self-evident.

Stochastic mechanics can be regarded as an effective theory for describing a 
complete mechanical system within specific spatial and temporal 
scales \cite{ford1965statistical,mori1965transport,zwanzig1973nonlinear}.
Let us consider a heavy particle in the above system, assuming that the 
remaining particles constitute a heat reservoir that has already reached 
equilibrium. The electromagnetic interaction acting on the heavy particle 
can be divided into two parts: the coarse-grained averaging effects at larger 
spatial and temporal scales, and the stochastic remnants at smaller scales. 
Consequently, eqs.~\eqref{5.33}-\eqref{5.34} can be approximated by LE$_\tau$ 
\cite{cai2023relativistic}, and this heavy particle is the Brownian particle.

However, such a coarse-grained description violates the TRS. Let us first 
consider the damping force. If the forward and reversed trajectories 
simultaneously satisfy the following equation
\begin{align}\label{damping-eq}
\frac{1}{m}p^\nu\nabla_\nu p^\mu=\mathcal K^{\mu\nu}U_\nu,
\end{align}
the damping tensor must reverse its sign under the TRT, since the velocity of 
the heat reservoir should reverse its sign. The sign change in $\mathcal K^{\mu\nu}$ implies 
that the damping force reverses its role from a decelerating force to an accelerating 
force, which makes a difference between future- and past-directed observers. 
On the other hand, the Einstein relation \cite{cai2023relativistic2}
\begin{align}
\mathcal{D}^{\mu\nu}=2T_{\rm B}\mathcal{K}^{\mu\nu}
\end{align}
requires that the damping tensor should be kept invariant under the TRT, 
because the diffusion tensor $\mathcal D^{\mu\nu}
=\mathcal{R}\ind{\mu}{\fa}\mathcal R\ind{\nu}{\fa}$ is a quadratic form 
in $\mathcal{R}\ind{\mu}{\fa}$ whose eigenvalues must be non-negative. 
Therefore, the assumption that eq.\eqref{damping-eq} is invariant under the TRT 
has to be wrong. The correct behaviors of the damping and diffusion tensors 
and the stochastic amplitudes are provided as follows,
\begin{align}
I^*\mathcal{K}^{\mu\nu}=\mathcal{K}^{\mu\nu},\qquad 
I^*\mathcal{D}^{\mu\nu}=\mathcal{D}^{\mu\nu},\qquad 
I^*\mathcal{R}\ind{\mu}{\fa}= \pm \mathcal{R}\ind{\mu}{\fa}.
\end{align}
The sign of the stochastic amplitude $\mathcal{R}\ind{\mu}{\fa}$ is of little 
significance, since the probability distribution of the Gaussian noise is an even 
function. For convenience, we adopt $I^*\mathcal{R}\ind{\mu}{\fa}=  
\mathcal{R}\ind{\mu}{\fa}$.
For the sake of mathematical consistency, we define a vector field $V^a$ on 
$T\mathcal M$, which satisfies
\begin{align}
V^a|_{(x,p)}=\begin{cases}
\displaystyle U^\mu|_x\pfrac{}{p^\mu}\bigg|_{(x,p)} &(x,p)\in \Gamma^+_m\\
\displaystyle -U^\mu|_x\pfrac{}{p^\mu}\bigg|_{(x,p)} &(x,p)\in \Gamma^-_m
\end{cases}.
\end{align}
Then the damping force and the right-hand side of eq.~\eqref{Liouville-II} 
can be rewritten as 
\begin{align}
\mathcal F_{\text{dp}}^a=\mathcal{K}^{ab}V_{b},\qquad 
\mathcal I^a[\varphi]=\frac{1}{2}\mathcal{D}^{ab} \nabla^{(h)}_{b}\varphi
-\mathcal{K}^{ab}V_{b}\varphi,
\end{align}
and it is easy to check that their transformation rules under the TRT are
\begin{align}
I^*\mathcal{F}_{\text{dp}}=\mathcal{F}_{\text{dp}},\qquad 
I^*\mathcal{I}[\varphi]=\mathcal{I}[\varphi].
\end{align}
Therefore, the probability current \eqref{probability-current} of the 
Brownian particle can be divided into the even and odd parts under the TRT, \ie
\begin{align}
\mathscr{J}_{\text{r}}[\varphi]=\dfrac{\varphi}{m}\mathscr{L}_{F},\qquad
\mathscr{J}_{\text{d}}[\varphi]=-\mathcal{I}[\varphi], 
\end{align}
with
\begin{align}
I^*\mathscr{J}_{\text{r}}[\varphi]=-\mathscr{J}_{\text{r}}[\varphi],\qquad 
I^*\mathscr{J}_{\text{d}}[\varphi]=\mathscr{J}_{\text{d}}[\varphi].
\end{align}
It is evident that $\mathscr{J}_{\text{d}}[\varphi]$ violates the TRS. 
The entropy production is always closely related to the breaking of the TRS, 
and we have proved \cite{wang2023general} that, on the ensemble level, 
\begin{align}
&\nabla_{\mu}S^{\mu}[\varphi]=-\int\eta_{(\Gamma_{m}^{+})_{x}}
\varphi^{-1}\pfrac{\varphi}{p^\mu}\mathscr{J}^\mu_{\mathrm{d}}[\varphi],\\
&\nabla_{\mu}S_{R}^{\mu}=\int\eta_{(\Gamma_{m}^{+})_{x}}\beta_{\mu}
\mathscr{J}^{\mu}_{\text{d}}[\varphi],
\label{nabSr}
\end{align}
where $S^{\mu}[\varphi]$ and $S_{R}^{\mu}$ respectively denote the entropy currents 
of the Brownian particle and of the heat reservoir. 
Eq.~\eqref{nabSr} is actually the relativistic version of the celebrated Clausius' identity
\begin{align}
\nabla_\mu S^\mu_{R}=-\frac{Q_{\rm B}[\varphi]}{T_{\rm B}}=\frac{Q_R}{T_{\rm B}},
\end{align}
where $Q_{\rm B}[\varphi]$ is the heat transfer rate  
from Bob's perspective, and $Q_R$ is that of the heat reservoir. 
It is worth noticing that Clausius' identity holds only from the perspective of Bob.

\section{Fluctuation theorem}
\label{Sec:fluctuation}

Unlike the case of ordinary differential equations, there is no deterministic 
solution for stochastic differential equation. The best one can do is to 
determine the probability for a certain trajectory to be realized. Consequently, 
the breaking of the TRS of the Langevin equation could be described 
in terms of the non-equal probabilities for the 
forward and reversed trajectories,
\begin{align}
\Pr[\tilde Y_{[t]}=Y_{[t]}]\neq \Pr[\tilde Y_{[t]}^-=Y_{[t]}^-],
\end{align}
where $\tilde Y_{[t]}=(\tilde Y_0,\tilde Y_1,\cdots,\tilde Y_N)$ and 
$ \tilde Y_{[t]}^-=(\tilde Y^-_0,\tilde Y_1^-,\cdots,\tilde Y_N^-)$ denote 
the forward and reversed processes, 
and $Y_{[t]}$ and $ Y^-_{[t]}$ denote the  forward and reversed trajectories 
which are already described in Sec.~\ref{Sec:time-reversal}. Notice that  
``forward process'' and ``forward trajectory'' are different concepts: 
the latter is a concrete realization of the former. 
Unlike the relation between forward and reversed 
trajectories, the only requirement in the reversed process is that its initial state 
is identical to the TRT of the final state of the forward process, \ie
\begin{align}\label{initial-final-probability}
\tilde Y^-_0= I(\tilde Y_N).
\end{align}

\blue{Using the above conventions, we are now in the right 
position to present the precise form of the detailed and integral fluctuation theorems 
in the context of general relativistic stochastic thermodynamics 
and formulate their proofs.}

\blue{
\begin{theorem}[Detailed fluctuation theorem]\label{thm5.1}
The ratio between the probabilities for the forward and reversed 
general relativistic stochastic trajectories $Y_{[t]}$ and $Y^-_{[t]}$ 
to be realized is equal to the exponential of total entropy production $\Sigma_{Y_{[t]}}$
along the trajectory $Y_{[t]}$, \ie
\begin{align}\label{detialed-fluctuation}
\frac{\Pr[\tilde Y_{[t]}=Y_{[t]}]}{\Pr[\tilde Y^-_{[t]}=Y^-_{[t]}]}=\re^{\Sigma_{Y_{[t]}}}.
\end{align}
\end{theorem}
}

\blue{
\begin{theorem}[Integral fluctuation theorem]\label{thm5.2}
The statistical expectation value of the total trajectory entropy production 
must be non-negative, \ie
\begin{align}
\re^{-\left\langle \Sigma_{Y_{[t]}}\right\rangle}\leq 1,\qquad
\left\langle \Sigma_{Y_{[t]}}\right\rangle \geq 0.
\end{align}
\end{theorem}
}

\blue{The rest of this section is devoted to the proof of the above theorems.}

The Lorentz factors relative to Alice and Carol can be related via the TRT
\begin{align}
\gamma_{\rm A}=-\lambda Z_\mu p^\mu/m=\lambda C_\mu p^\mu/m=I^*\gamma_C.
\end{align}
These Lorentz factors can be merged into a single scalar field on the complete 
mass shell bundle,
\begin{align}
\gamma:=\begin{cases}
\gamma_{\rm A}|_Y&Y\in \Gamma^+_m\\
\gamma_C|_Y&Y\in \Gamma^-_m
\end{cases},
\end{align}
which is even under the TRT, $I^*\gamma=\gamma$. 
Accounting for the transformation rules under the TRT,
we can rearrange the spatial components of the LE$_t$ presented in 
eqs.~\eqref{38}-\eqref{39} into the form
\begin{align}\label{LE_Alice_1}
\rd \tilde y_t^i&=\frac{\tilde{k}^i_t}{m}\gamma^{-1}\rd t,\\
\label{LE_Alice_2}
\rd \tilde k_t^i&=\hat{\mathcal{R}}\ind{i}{\fa}\circ_S\rd \tilde W^\fa_n
+F^i\rd t+\bar{F}^i\rd t,
\end{align}
where {$\hat{\mathcal{R}}\ind{i}{\fa}:=\gamma^{-1/2}\mathcal{R}\ind{i}{\fa}$} and
\begin{align}
F^i:=\gamma^{-1}\(\mathcal F^i_{\text{em}}-\frac{1}{m}\varGamma\ind{i}
{\alpha\beta}\tilde k^\alpha \tilde k^\beta \),\quad
\bar F^i&:=\gamma^{-1}\(\mathcal{F}_\add^i
-\frac{1}{2}\mathcal{D}^{ij}\gamma^{1/2}\nabla^{(h)}_j\gamma^{-1/2}
+\mathcal{F}^i_{\text{dp}} \),
\end{align}
which have opposite behaviors under the TRT,
\begin{align}
F^i|_{Y}=F^i|_{I(Y)}, \qquad \bar F^i|_{Y}=-\bar F^i|_{I(Y)}.
\end{align}

In order to \blue{prove} the fluctuation theorems at the trajectory level, 
we need a discretized version of LE$_t$, which reads
\begin{align}\label{LE_Alice_1}
\rd \tilde y_n^i&=\frac{\tilde{k}^i_n}{m}\gamma^{-1}|_{\tilde Y_{\bar n}}\rd t,\\
\label{LE_Alice_2}
\rd \tilde k_n^i&=\hat{\mathcal{R}}\ind{i}{\fa}|_{\tilde Y_{\bar n}} \rd \tilde W^\fa_n
+F^i|_{\tilde Y_{\bar n}}\rd t+\bar{F}^i|_{\tilde Y_{\bar n}}\rd t,
\end{align}
where $\tilde Y_{\bar n}:=(\tilde Y_{n+1}+\tilde Y_n)/2$ which comes from the Stratonovich
coupling. We also need to introduce the measures on the space of 
forward and reversed trajectories, 
which are defined to be the successive wedge products of the volume element of the 
space of microstates at each moment of the discrete time, 
\begin{align}
\mathcal{D}[Y_{[t]}]:=\eta_{\Sigma_0}\wedge\eta_{\Sigma_1}\wedge\cdots
\wedge\eta_{\Sigma_N},\qquad
\mathcal{D}[Y^-_{[t]}]:=\eta_{\Sigma^-_0}\wedge
\eta_{\Sigma^-_1}\wedge\cdots\wedge\eta_{\Sigma^-_N}.
\end{align}
Eq.~\eqref{TRT-volume} implies that these measures are images of each other under the TRT,
\begin{align}
I^*\mathcal{D}[Y_{[t]}]=\mathcal{D}[Y^-_{[t]}].
\end{align}
In other words, $Y_{[t]}\mapsto Y^-_{[t]}$ is a volume preserving map.

Since both the forward and reversed processes are Markovian, the trajectory probabilities 
can be written as products of transition probabilities with the initial probability,
\begin{align}
\Pr[\tilde Y_{[t]}=Y_{[t]}]&=
\bigg(\prod_{n=0}^{N-1}\Pr[\tilde Y_{n+1}=Y_{n+1}|\tilde Y_n=Y_n]
\bigg) \Pr[\tilde Y_0=Y_0],\\
\Pr[\tilde Y^-_{[\bar t\,]}=Y^-_{[t]}]&=
\bigg(\prod_{n=0}^{N-1}\Pr[\tilde Y^-_{n+1}=Y^-_{n+1}|\tilde Y^-_n=Y^-_n]
\bigg)\Pr[\tilde Y^-_0=Y^-_0]\notag\\
&=\bigg(
\prod_{n=0}^{N-1}\Pr[\tilde Y^-_{n+1}=I(Y_{N-n-1})|\tilde Y^-_n=I(Y_{N-n})]
\bigg)\Pr[I(\tilde Y_N)=I(Y_N)]\notag\\
&=\bigg(\prod_{n=0}^{N-1}\Pr[\tilde Y^-_{N-n}=I(Y_{n})|\tilde Y^-_{N-n-1}=I(Y_{n+1})]
\bigg)\Pr[\tilde Y_N=Y_N].
\label{probability-reversal-trajectory}
\end{align}
Eq.~\eqref{relation-trajectory} and eq.~\eqref{initial-final-probability} are used 
in the second line of eq.~\eqref{probability-reversal-trajectory}, and the 
volume preserving property of $I$ is used in the third line. 
Appendix~\ref{Sec:limit} proves the following continuum limit 
\begin{align}\label{continuum-limit}
&\lim_{N\rightarrow +\infty}\prod_{n=0}^{N-1}
\frac{\Pr[\tilde Y_{n+1}=Y_{n+1}|\tilde Y_n=Y_n]}
{\Pr[\tilde Y^-_{N-n}=I(Y_{n})|\tilde Y^-_{N-n-1}=I(Y_{n+1})]}\notag\\
&\qquad =\frac{(\lambda\gamma^{-1})|_{Y_0}}{(\lambda\gamma^{-1})|_{Y_N}}
\exp\left[\int^{t_F}_{t_I}\gamma^{-1}\rd t\frac{1}{T_{\rm B}}
({m} a^\mu-\mathcal{F}^\mu_{\text{em}})U_\mu \right].
\end{align}
Under the perspective of Bob, the complete differential of the energy is
\begin{align}
\rd E_{\rm B}=-\frac{p^\nu}{m}\nabla_\nu(p^\mu U_\mu)\rd \tau
=-ma^\mu U_\mu\rd \tau+\rd \mathcal{P},
\end{align}
or
\begin{align}
\rd H_{\rm B}=-\frac{p^\nu}{m}\nabla_\nu(P^\mu U_\mu)\rd \tau
=-ma^\mu U_\mu\rd \tau+\rd \mathcal{P}+{\rd_\sigma\mathcal{U}}-\rd \mathcal W.
\end{align}
Comparing with eq.~\eqref{dE}, it is clear that the integral on the exponent 
in eq.~\eqref{continuum-limit} is actually the increase of the entropy of the heat 
reservoir
\begin{align}\label{Clausius}
\int\gamma^{-1}\rd t\frac{1}{T_{\rm B}}(m a^\mu-\mathcal{F}^\mu_{\text{em}})U_\mu
=-\int\frac{\rd \mathcal Q}{T_{\rm B}}=\Delta S_R,
\end{align}
where the second equality encodes the Clausius equality at the level of trajectories. 
Let us stress that, although the Clausius equality holds only from the perspective of 
Bob, the total increase of the entropy of the heat reservoir is actually 
observer-independent. This fact enables us to make use of eq.~\eqref{Clausius} in 
addressing the fluctuation theorem from the perspective of an arbitrary observer.

In the continuum limit, the logarithm of the ratio of the probabilities of 
the forward and reversed trajectories is the sum of the trajectory entropy production of 
the Brownian particle with the entropy increase of the heat reservoir,
\begin{align}
\ln\frac{\Pr[\tilde Y_{[t]}=Y_{[t]}]}{\Pr[\tilde Y^-_{[t]}=Y^-_{[t]}]}&=
\lim_{N\to\infty} \sum_{n=0}^{N-1} \ln\frac{\Pr[\tilde Y_{n+1}=Y_{n+1}|\tilde Y_n=Y_n]}
{\Pr[\tilde Y^-_{N-n}=I(Y_{n})|\tilde Y^-_{N-n-1}=I(Y_{n+1})]}\nonumber\\
&\qquad +\ln\frac{(\lambda\gamma^{-1})|_{X_0}\Pr[\tilde Y_0=Y_0]}
{(\lambda\gamma^{-1})|_{Y_N}\Pr[\tilde Y_N=Y_N]}\notag\\
&=-\int\frac{\rd \mathcal Q}{T_{\rm B}}
+\ln\frac{\varphi(Y_0)}{\varphi(Y_N)}\notag\\
&=\Delta S_R+\Delta S
\blue{~:= \Sigma_{Y_{[t]}}}.
\end{align}
It is important to remember that the relation eq.~\eqref{pro-microstate} between 
different distribution functions and the definition \eqref{trajectory-entropy} of 
the trajectory entropy production are used here. 
\blue{Exponentiating the above relation yields Theorem \ref{thm5.1}.
Moreover, Theorem \ref{thm5.2} follows immediately by use of Theorem \ref{thm5.1}
and taking an integration in the space of trajectories, together with the aid of 
Jensen inequality, 
\begin{align}
\re^{-\left\langle \Sigma_{Y_{[t]}}\right\rangle}\leq 
\left\langle \re^{-\Sigma_{Y_{[t]}}} \right\rangle
&:=\int\mathcal{D}[Y_{[t]}]\Pr[\tilde Y_{[t]}=Y_{[t]}]~ \re^{-\Sigma_{Y_{[t]}}}\notag\\
&=\int\mathcal{D}[Y^-_{[t]}]\Pr[\tilde Y^-_{[t]}=Y^-_{[t]}]=1,
\end{align}
which implies $\left\langle \Sigma_{Y_{[t]}}\right\rangle\geq 0$.
}

\section{Concluding remarks}
\label{Sec:conclusion}

\blue{We have thus proved the detailed and integral fluctuation theorems in the context of 
general relativistic stochastic thermodynamics. }
In our construction, it is important to decouple the choice of observers from
the choice of coordinate systems, which has several important consequences.
First, it allows us to parametrize the stochastic trajectory using the 
proper time of the observer field, which helps to get rid of the 
random clock carried by the Brownian particle. Second, it helps to 
properly describe the TRT in a coordinate independent manner. Lastly,
it is precisely such decoupling that makes the 
construction fully general covariant. 

Although the values of 
many thermodynamic quantities are observer dependent, including \eg 
the energy and its density, the temperature and chemical potential, 
and even the Clausius' identity {\em etc.}, 
the total entropy production is not among the observer dependent quantities. 
Meanwhile, the trajectory probability is a pure mathematical entity
that is also observer independent. Therefore, it is not surprising that
the form of the fluctuation theorems obtained in this work is identical to that 
obtained from non-relativistic stochastic thermodynamics. 

Fluctuation theorems are not only important in understanding the theoretical origin of 
irreversibility, but they have also found various applications 
in diverse areas ranging from macroscopic to mesoscopic systems, and even to
active matter. \blue{For example, there are some studies 
\cite{qian2021new,mandal2012work,esposito2012stochastic,hartich2014stochastic} 
utilizing the non-relativistic fluctuation theorem to explore the concept of 
Maxwell's demon. Some of them argue that the change of the mutual information 
between Maxwell's demon and the Brownian particle should be considered as a part of 
the entropy production. When considering the event horizon within curved spacetime, 
which serves as a causal boundary, the causal connection between Maxwell's demon and 
the Brownian particle may be lost over time. Consequently, behaviors different 
from those predicted by non-relativistic fluctuation theorems, similar to 
the black hole information paradox, may emerge. 
Another possible scenario in which our general relativistic fluctuation theorems 
may find application is in black hole physics. In recent years, due to the 
observational progresses brought about by the event horizon telescope \cite{EVH}, 
a considerable amount of theoretical works on the images and shadows of black holes
have appeared\footnote{A quick search of the literature \cite{inspirehep} has 
found 3222 results. It is impossible to list them all, therefore we simply refer to 
\cite{GHW,perlick2022calculating,Lima} as prototypes of related works.}. 
In most of those works, the light-emitting substance is assumed 
to be particles undergoing regular geodesic motion, rendering the corresponding images
also regular and containing rich patterns. In reality, however, 
since a large amount of particles accrete around the black holes, they effectively form 
a heat reservoir with a very high temperature, and consequently they make the trajectories of 
individual particles moving inside the accretion disk probabilistic. The corresponding 
images should also be obscured as a consequence of such probabilistic motion. 
The quantitative nature of the detailed fluctuation theorem may help to understand 
to what extent the black hole images should be obscured. In summary, 
} 
the research carried out in the present work opens a new area 
for the potential applications of fluctuation theorems which embodies 
relativistic gravity, as in typical cases of cosmological processes, astrophysical 
processes and \blue{black hole physics. We believe that much more sophisticated 
applications of the new version of the fluctuation theorems presented in this work and 
more generally of the framework of general relativistic stochastic thermodynamics 
will emerge as we look more clearly and deeply into them. We are working hard on 
related problems.}

\section*{Acknowledhement} 

This work is supported by the National Natural Science 
Foundation of China under the grant No. 12275138.

\section*{Conflict of interests declaration}  

The authors declare no known conflict of interests.

\begin{appendices}

\section{Coordinate bases for tangent vectors}
\label{app:tangent-vector}

In our description of the tangent bundle $T\mathcal{M}$, we employ $2d+2$ 
independent coordinates $(x^\mu,p^\mu)$ with the corresponding 
coordinate basis $\displaystyle
\left(\left.\frac{\partial}{\partial x^\mu}\right|_{T\mathcal{M}},
\left.\frac{\partial}{\partial p^\mu}\right|_{T\mathcal{M}} \right)$ 
for tangent vectors on $T\mathcal{M}$. 
However, due to the mass shell constraint, the above coordinates are no longer 
independent of the mass shell bundle $\Gamma^+_m$, 
rendering the coordinate basis also redundant. The same also happens when 
we restrict ourselves from the mass shell bundle to the 
space of microstates $\Sigma_t^+$. The aim of this appendix is to 
resolve the redundancies in the coordinate bases on various useful submanifolds 
of $T\mathcal{M}$.

To resolve the coordinate redundancy on $\Gamma^+_m$, we take the coordinates 
on $\Gamma^+_m$ to be $(x^\mu,p^i)$ and view $p^0$ as a function 
\begin{align}
p^0(x^\mu,p^i) = \frac{g_{0i}(x)p^i \pm \sqrt{\left[g_{0i}(x)p^i\right]^2 
- g_{00}(x)\left[m^2 + g_{ij}(x)p^i p^j\right]}}{-g_{00}(x)}.
\end{align}
This leads to the partial derivatives
\begin{align}\label{partial-p0}
\frac{\partial p^0}{\partial x^\mu} 
= -\frac{1}{2p_0}\frac{\partial g_{\alpha\beta}}{\partial x^\mu}p^\alpha p^\beta, \qquad
\frac{\partial p^0}{\partial p^i} = -\frac{p_i}{p_0}.
\end{align}
It is essential to recognize that the coordinate basis for tangent vectors on 
$\Gamma^+_m$ is different from the corresponding subset of coordinate basis 
on $T\mathcal{M}$,
\begin{align}
\left.\frac{\partial}{\partial x^\mu}\right|_{T\mathcal{M}} \neq 
\left.\frac{\partial}{\partial x^\mu}\right|_{\Gamma^+_m}, \quad 
\left.\frac{\partial}{\partial p^i}\right|_{T\mathcal{M}} \neq 
\left.\frac{\partial}{\partial p^i}\right|_{\Gamma^+_m}.
\end{align}
To demonstrate this difference, let us consider a scalar field $f(x^\mu,p^\mu)$ on 
$T\mathcal M$ and its restriction on $\Gamma^+_m$, \ie
\begin{align}\label{def_restriction}
\left.f\right|_{\Gamma^+_m}(x^\mu,p^i):=f(x^\mu,p^0(x^\mu,p^i),p^i).
\end{align}
The action of the coordinate basis vectors on $\left.f\right|_{\Gamma^+_m}$ reads
\begin{align}
\label{relation-partial1}
\left.\pfrac{}{p^i}\right|_{\Gamma^+_m}\left.f\right|_{\Gamma^+_m}
&=\left.\pfrac{}{p^i}\right|_{T\mathcal{M}}f+\pfrac{p^0}{p^i}
\left.\pfrac{}{p^0}\right|_{T\mathcal{M}}f
=\left.\pfrac{}{\breve{p}^i}\right|_{T\mathcal{M}}f, \\
\label{relation-partial2}
\left.\pfrac{}{x^\mu}\right|_{\Gamma^+_m}\left.f\right|_{\Gamma^+_m}
&=\left.\pfrac{}{x^\mu}\right|_{T\mathcal{M}}f+\pfrac{p^0}{x^\mu}
\left.\pfrac{}{p^0}\right|_{T\mathcal{M}}f
=\left.\pfrac{}{\breve{x}^\mu}\right|_{T\mathcal{M}}f,
\end{align}
where the ``breved'' partial derivatives are defined as
\begin{align}
\left.\pfrac{}{\breve{p}^i}\right|_{T\mathcal{M}}
:=\left[\pfrac{}{p^i}-\frac{p_i}{p_0}\pfrac{}{p^0} \right]_{T\mathcal{M}},\qquad 
\left.\pfrac{}{\breve{x}^\mu}\right|_{T\mathcal{M}}
:=\left[e_\mu+\varGamma\ind{i}{\beta\mu}p^\beta\pfrac{}{\breve p^i} \right]_{T\mathcal{M}}.
\end{align}
Therefore, the partial derivatives on $\Gamma^+_m$ and $T\mathcal M$ are related via
\begin{align}
i^*\left.\pfrac{}{p^i}\right|_{\Gamma^+_m}
=\left.\pfrac{}{\breve{p}^i}\right|_{T\mathcal{M}},\qquad 
i^*\left.\pfrac{}{x^\mu}\right|_{\Gamma^+_m}
=\left.\pfrac{}{\breve{x}^\mu}\right|_{T\mathcal{M}},
\end{align}
where $i:\Gamma^+_m\rightarrow T\mathcal{M}$ is the embedding map. 
From the exterior geometric point of view, we can safely identify 
$\{\partial/\partial \breve x^\mu,\partial/\partial \breve p^i \}$ 
as the coordinate basis on the mass shell bundle, omitting the subscripts.

The tangent vector $\mathcal{V}=\mathcal{V}^\mu(\partial/{\partial p^\mu})$ 
of the momentum space must be orthogonal to the normal vector 
$\hat{N}=p^\mu(\partial/\partial p^\mu)/m$. The orthogonality 
condition $\mathcal{V}^\mu p_\mu=0$ implies that
\begin{align}
\mathcal V^\mu \pfrac{}{p^\mu}
=\mathcal{V}^i\pfrac{}{p^i}-\frac{\mathcal V^i p_i}{p_0}\pfrac{}{p^0}
=\mathcal V^i\pfrac{}{\breve p^i}.
\end{align}
Therefore, the tangent vectors of the momentum space have two different 
representations under the above two bases. We will use them interchangeably.
There is a similar property for the vectors on the 
mass shell bundle. Let $\mathscr{V}$ be a tangent vector of the phase 
trajectory $X_\tau:=(x_\tau, p_\tau)$. 
As a vector on a tangent bundle, $\mathscr{V}$ can be expanded as
\begin{align}
\mathscr{V}=\frac{p^\mu}{m}\pfrac{}{x^\mu}+m\mathcal{A}^\mu\pfrac{}{p^\mu}
=\frac{p^\mu}{m}e_\mu+ma^\mu\pfrac{}{p^\mu},
\end{align}
where $\mathcal{A}^\mu:=m^{-1}\rd p^\mu/\rd\tau$ is the {\em coordinate acceleration} and 
$a^\mu:=\mathcal{A}^\mu+\varGamma\ind{\mu}{\alpha\beta}p^\alpha p^\beta/m^2$ is the 
{\em covariant acceleration}. The mass shell condition implies that 
the covariant acceleration must always be orthogonal to the momentum, 
so that $\mathscr{V}$ can be rewritten as
\begin{align}\label{phase-trajectory-vector}
\mathscr{V}=\frac{p^\mu}{m}e_\mu+ma^i\pfrac{}{\breve p^i}
=\frac{p^\mu}{m}\pfrac{}{\breve x^\mu}+m\mathcal{A}^i\pfrac{}{\breve p^i}.
\end{align}
The two representations for vectors on the mass shell bundle under the 
bases $\{\partial/\partial\breve x^\mu,\partial/\partial \breve p^i \}$ 
and $\{\partial/\partial x^\mu,\partial/\partial  p^\mu \}$ 
will also be used interchangeably.

The space of microstates $\Sigma^+_t$ is also an embedding submanifold of 
the mass shell bundle defined by fixing $t(x)$ to be a constant. 
Therefore, the coordinates on $\Gamma^+_m$ become redundant once again on $\Sigma^+_t$. 
Such redundancy can also be eliminated by taking $(x^i,p^i)$ as a coordinate 
of $\Sigma^+_t$ and viewing $x^0(t,x^i)$ as a function. The restriction of $f$ 
on $\Sigma^+_t$ is defined as
\begin{align}
\left.f\right|_{\Sigma_t^+}(x^i,p^i)
:=\left.f\right|_{\Gamma^+_m}(x^0(t,x^i),x^i,p^i)
=f(x^0(t,x^i),x^i,p^0(x^0(t,x^i),x^i,p^i),p^i).
\end{align}
Correspondingly, the partial derivatives acting on $f|_{\Sigma^+_t}$ can be 
evaluated to be 
\begin{align}
\left.\pfrac{}{x^i}\right|_{\Sigma_t^+}\left.f\right|_{\Sigma_t^+}
&=\left.\pfrac{}{x^i}\right|_{\Gamma^+_m}\left.f\right|_{\Gamma^+_m}
+\pfrac{x^0}{x^i}\left.\pfrac{}{x^0}\right|_{\Gamma^+_m}\left.f\right|_{\Gamma^+_m}
=\left.\pfrac{}{\breve{x}^i}\right|_{T\mathcal{M}}f
-\frac{\partial_i t}{\partial_0 t}\left.\pfrac{}{\breve{x}^0}\right|_{T\mathcal{M}}f,\\
\left.\pfrac{}{p^i}\right|_{\Sigma_t^+}\left.f\right|_{\Sigma^+_t}
&=\left.\pfrac{}{p^i}\right|_{\Gamma^+_m}\left.f\right|_{\Gamma^+_m}
=\left.\pfrac{}{\breve{p}^i}\right|_{T\mathcal{M}}f,\\
\label{def-parrtialt}
\pfrac{}{t}\left.f\right|_{\Sigma^+_t}
&=\pfrac{x^0}{ t}\left.\pfrac{}{x^0}\right|_{\Gamma^+_m}\left.f\right|_{\Gamma^+_m}
=\frac{1}{\partial_0 t}\left.\pfrac{}{\breve x^0}\right|_{T\mathcal{M}}f,
\end{align}
thanks to the relations \eqref{relation-partial1}-\eqref{relation-partial2}. 
Let $\pi_t:\Sigma_t^+\rightarrow\Gamma^+_m$ be the embedding map and define
\begin{align}
\left.\pfrac{}{\hat x^i}\right|_{T\mathcal M}
:=\left.\pfrac{}{\breve x^i}\right|_{T\mathcal M}
-\frac{\partial_i t}{\partial_0 t}\left.\pfrac{}{\breve x^0}\right|_{T\mathcal M},
\end{align}
we have
\begin{align}
(i\circ\pi_t)^* \left.\pfrac{}{p^i}\right|_{\Sigma_t^+}
=\left.\pfrac{}{\breve p^i}\right|_{T\mathcal M},\qquad 
(i\circ\pi_t)^* \left.\pfrac{}{x^i}\right|_{\Sigma_t^+}
=\left.\pfrac{}{\hat x^i}\right|_{T\mathcal M}.
\end{align}
Therefore, we can take $\{\partial/\partial \hat x^i,\partial/\partial \breve p^i\}$
as the coordinate basis on $\Sigma^+_t$ without specifying the manifold with a subscript.

Since the phase trajectory does not lie on a certain $\Sigma^+_t$, the basis
$\{\partial/\partial \hat x^i,\partial/\partial \breve p^i\}$ is insufficient to
describe the tangent vector of the phase trajectory. The missing dimension in
the tangent vector of the phase trajectory is described by $\partial/\partial t$ 
as described in eq.~\eqref{def-parrtialt}, which, from the exterior geometric 
point of view, can be simply denoted
\begin{align}
\pfrac{}{t}=\frac{1}{\partial_0 t}\pfrac{}{\breve x^0}.
\end{align}
Therefore, under the above vector basis, the tangent vector 
\eqref{phase-trajectory-vector} to the phase trajectory can be re-expressed as
\begin{align}\label{phase-trajectory-vector-2}
\mathscr{V}=\gamma\pfrac{}{t}+\frac{p^i}{m}\pfrac{}{\hat{x}^i}
+m \mathcal A^i\pfrac{}{\breve p^i},
\end{align}
where 
\begin{align}
\gamma = \frac{\rd t}{\rd \tau}=\frac{1}{m}p^\mu\partial_\mu t 
\label{ttau}
\end{align}
is the local Lorentz factor between the particle and the observer Alice.

\section{Continuum limit}
\label{Sec:limit}

This appendix provides the details for proving eq.~\eqref{continuum-limit}. 

First, we introduce two mathematical lemmas. 

1. Let $A$ be a full rank square matrix, and let $B$ be an arbitrary matrix
of the same size. Then the determinant of $A+B\rd t+o(t^{2})$ can be expanded into
power series in $\rd t$,
\begin{align}
\det[A+B\rd t+o(\rd t^{2})]=\det[A]+\det[A]\mathrm{Tr}[A^{-1}B]\rd t+o(\rd t^{2}).
\end{align}

2. Let $f(t)$ be a continuous function on $[t_{I},t_{F}]$, and then 
\begin{align}
    \lim_{N\rightarrow +\infty}\prod_{n=0}^{N-1}[1+f(t_{n})\rd t+o(\rd t^{2})]=\exp\left[\int_{t_{I}}^{t_{F}}f(t) \rd t \right],
\end{align}
where $\rd t=(t_{F}-t_{I})/N$ and $t_{n}\in[t_I+n\rd t,t_I+(n+1)\rd t]$. 

Combining the above two lemmas, the following corollary can be deduced,
\begin{align}\label{lamme-continuum-limite}
\lim_{N\rightarrow +\infty}\prod_{n=0}^{N-1}\frac{\det[A(t_{n})+B(t_{n})\rd t
+o(\rd t^{2})]}{\det[A(t_{n})+C(t_{n})\rd t+o(\rd t^{2})]}
=\exp\left[\int_{t_{I}}^{t_{F}}\mathrm{Tr}[A^{-1}(B-C)]\rd t \right],
\end{align}
where $A(t)$, $B(t)$ and $C(t)$ are all time-dependent matrix functions.

Let $\mathscr{O}$ be a function on the mass shell bundle, and let $Y_t =X_{\tau(t)}$ be a 
phase trajectory parametrized by the proper time of Alice. Then 
\begin{align}\label{ratio-f-i}
\frac{\mathscr{O}(Y_{t_{F}})}{\mathscr{O}(Y_{t_{I}})}&=
\frac{\mathscr{O}(X_{\tau_{F}})}{\mathscr{O}(X_{\tau_{I}})}
=\lim_{N\rightarrow+\infty}\prod_{n=0}^{N-1}
\frac{\mathscr{O}(X_{\tau_{n+1}})}{\mathscr{O}(X_{\tau_{n}})}\notag\\
&=\lim_{N\rightarrow+\infty}\prod_{n=0}^{N-1}\frac{\mathscr{O}(X_{\tau_{n}})
+\mathscr{V}(\mathscr{O})\rd \tau+o(\rd \tau^{2})}{\mathscr{O}(X_{\tau_{n}})}\notag\\
&=\exp\left[\int_{\tau_{I}}^{\tau_{F}}
\mathscr{O}^{-1}\mathscr{V}(\mathscr{O})\rd \tau\right]\notag\\
&=\exp\left[\int_{t_{I}}^{t_{F}}\gamma^{-1}\mathscr{O}^{-1}
\mathscr{V}(\mathscr{O})\rd t\right],
\end{align}
where $\mathscr{V}$ is the tangent vector along $Y_{t}$, 
as presented by eq.~\eqref{phase-trajectory-vector} and/or 
eq.~\eqref{phase-trajectory-vector-2}.
It is worth noting that in this appendix, we will use $(y,k)$ to 
represent the coordinates of the tangent bundle, rather than $(x,p)$. 
This choice does not imply a coordinate transformation. It simply emphasizes 
that the discretized time corresponds to Alice's proper time $t$.

Eq.~\eqref{ratio-f-i} can be used to evaluate the continuum limit of certain
ratios which are useful in proving the fluctuation theorem. For instance, 
\begin{align}\label{rate-p0}
\frac{k_{0}|_{Y_{N}}}{k_{0}|_{Y_{0}}}
&=\exp\left[\int_{t_{I}}^{t_{F}}\frac{\gamma^{-1}}{k_{0}}
\left\{\frac{k^{\mu}}{m}\pfrac{}{y^{\mu}}k_{0}
+m \mathcal{A}^{\mu}\pfrac{}{k^{\mu}}k_{0} \right\}\rd t \right]\notag\\
&=\exp\left[\int_{t_{I}}^{t_{F}}\frac{\gamma^{-1}}{k_{0}}
\left\{\frac{1}{m}\partial_{\mu} g_{0\nu} k^{\mu} k^{\nu}
+m \mathcal{A}_{0}\right\}\rd t \right]\notag\\
&=\exp\left[\int_{t_{I}}^{t_{F}}\frac{\gamma^{-1}}{k_{0}}
\left\{\frac{1}{2m}\partial_{0} g_{\mu\nu} k^{\mu} k^{\nu}+m a_{0}\right\}\rd t \right],
\\
\label{rate-g}
\frac{g|_{Y_{0}}}{g|_{Y_{N}}}&=
\exp\left[-\int_{t_{I}}^{t_{F}}\gamma^{-1}g^{-1} 
\frac{k^{\mu}}{m}\pfrac{}{y^{\mu}} g\rd t\right],
\end{align}
and
\begin{align}\label{rate-t0}
\frac{\partial_{0} t|_{Y_{N}}}{\partial_{0} t|_{Y_{0}}}
&=\exp\left[\int_{t_{I}}^{t_{F}}\frac{\gamma^{-1}}{\partial_{0} t}
\frac{k^{\mu}}{m}\pfrac{}{\breve{y}^{\mu}}\pfrac{}{\breve{y}^{0}}t \rd t \right]\notag\\
&=\exp\left[\int_{t_{I}}^{t_{F}}\frac{1}{\partial_{0} t}
\left\{\pfrac{}{\breve{y}^{0}}\(\gamma^{-1}\frac{k^{\mu}}{m}\partial_{\mu} t\)
-\partial_{\mu} t\pfrac{}{\breve{y}^{0}}\( \gamma^{-1}\frac{k^{\mu}}{m}\)\right\} 
\rd t \right]\notag\\
&=\exp\left[\int_{t_{I}}^{t_{F}}\frac{1}{\partial_{0} t}
\left\{\gamma^{-2}\frac{k^{\mu}}{m} \partial_{\mu} t\pfrac{}{\breve{y}^{0}} \gamma
-\frac{\gamma^{-1}}{m}\partial_{0} t\pfrac{}{\breve{y}^{0}}k^{0} \right\} \rd t \right]
\notag\\
&=\exp\left[\int_{t_{I}}^{t_{F}}\gamma^{-1}\pfrac{}{t}\gamma \rd t\right]
\exp\left[\int_{t_{I}}^{t_{F}}\frac{\gamma^{-1}}{k_{0}}
\frac{1}{2m}\partial_{0} g_{\mu\nu}k^{\mu} k^{\nu} \rd t\right].
\end{align}

Eq.~\eqref{LE_Alice_1} can be rewritten as
\begin{align}\label{def-h}
\tilde{y}^{i}_{n+1}=h^{i}(\tilde{Y}_{n+1},\tilde{Y}_{n})
=\frac{\tilde{k}^i_{\bar{n}}}{m}\gamma^{-1}|_{\tilde{Y}_{\bar{n}}}\rd t
+\tilde{y}^{i}_{n},
\end{align}
which implies that the realization of $\tilde{y}^{i}_{n+1}$ is determined by the 
realization of $\tilde{Y}_{n}$ and $\tilde{k}^{i}_{n+1}$. The probability 
distribution of a random variable $\tilde{x}$ obeying the constraint $\tilde{x}=f(\tilde{x})$ 
is given by
\begin{align}
\Pr[\tilde{x}=x]=(1-f'(x))\delta(x-f(x)).
\end{align}
The generalization of this equation to general dimension is straightforward, 
showing that the probability distribution of $\tilde{y}^{i}_{n+1}$ 
under the conditions $\tilde{k}^i_{n+1}={k}^i_{n+1},\tilde{Y}_{n}=Y_{n}$ is given by
\begin{align}\label{prob-y}
&\Pr[\tilde{y}_{n+1}=y_{n+1}|\tilde{k}_{n+1}={k}_{n+1},\tilde{Y}_{n}=Y_{n}]\notag\\
&\qquad=\det\left[\delta^{i}{}_{j}-\pfrac{h^{i}}{\hat{y}_{n+1}^{j}}\right]
\left|\frac{\partial_{0} t}{\lambda \sqrt{g}} \right|_{y_{n+1}}
\delta^{d}(y_{n+1}-h({Y_{n+1},Y_{n}}))\notag\\
&\qquad=\left[1+\gamma^{-2}\frac{k^{i}}{2m}\pfrac{}{\hat{{y}}^{i}}\gamma \rd t
+o(\rd t^{2})\right]_{Y_{\bar{n}}}\left|\frac{\partial_{0} t}
{\lambda \sqrt{g}} \right|_{y_{n+1}}\Delta(Y_{n+1},Y_{n}),
\end{align}
where $y_n$ denotes the sequence $(y_n^1,\cdots, y_n^d)$ and we will also use the notation
$k_n=(k^1_n,\cdots,k_n^d)$. $\Delta(Y_{n+1},Y_{n})$ is defined as 
$\Delta(Y_{n+1},Y_{n}):=\delta^{d}(y_{n+1}-h(Y_{n+1},Y_{n}))$, 
which has the property
\begin{align}
\Delta(Y_{n+1},Y_{n})=\Delta(I(Y_{n}),I(Y_{n+1})).
\end{align}
The appearance of $\partial/\partial\hat{y}^{i}$ in the second line of 
eq.~\eqref{prob-y} is due to the fact that $\tilde{y}^{i}_{n+1}$ only takes 
values in the configuration space, and its probability density is a scalar field 
on $\mathcal{S}_n$. The factor $|\partial_{0} t/(\lambda\sqrt{g})|$  arises from the change  
of coordinate volume element into an invariant volume element, 
as indicated by eq.~\eqref{volume-configuration}. 

The single step transition probability can be rewritten as
\begin{align}
\Pr[\tilde{Y}_{n+1}|\tilde{Y}_{n}]=\Pr[\tilde{y}_{n+1}|\tilde{k}_{n+1},\tilde{Y}_{n} ]
\Pr[\tilde{k}_{n+1}|\tilde{Y}_{n}],
\end{align}
where the second factor remains to be evaluated. 
Defining the function
\begin{align}
\rd W^{\mathfrak{a}}(Y_{n+1},Y_{n})
&:=(\hat{\mathcal{R}}^{-1})^{\mathfrak{a}}{}_{i}|_{Y_{\bar{n}}}
\left[k^{i}_{n+1}-k^{i}_{n}-F^{i}|_{Y_{\bar{n}}}\rd t
-\bar{F}^{i}|_{Y_{\bar{n}}}\rd t \right]\notag\\
&=(\hat{\mathcal{R}}^{-1})^{\mathfrak{a}}{}_{i}|_{Y_{\bar{n}}}
\left[\gamma^{-1}m\mathcal{A}^{i}_{n}-F^{i}-\bar{F}^{i} \right]_{Y_{\bar{n}}}\rd t,
\end{align}
where $\mathcal{A}^{i}_{n}:=\gamma|_{Y_{\bar{n}}}(k^{i}_{n+1}-k^{i}_{n})/(m\rd t)$ is 
the coordinate acceleration, eq.~\eqref{LE_Alice_2} can be rewritten as
\begin{align}\label{dw=dw}
\rd \tilde W^{\mathfrak{a}}_{n}=\rd W^{\mathfrak{a}}(\tilde{Y}_{n+1},\tilde{Y}_{n}).
\end{align}
In the case in which the realization of $\tilde{Y}_{n}$ is given, 
$\tilde{y}_{n+1}$ is determined by $\tilde{k}_{n+1}$. Therefore, 
eq.~\eqref{dw=dw} gives the relation between $\rd \tilde{W}^{\mathfrak{a}}_{n}$ 
and $\tilde{k}^i_{n+1}$, which can be used to calculate the conditional probability
\begin{align}\label{prod-p-X}
&\Pr[\tilde{k}_{n+1}=k_{n+1}|\tilde{Y}_{n}=Y_{n}]
=\left|\frac{k_{0}}{m\sqrt{g}}\right|_{Y_{n+1}}
\det\left[T^{\mathfrak{a}}{}_{i}(Y_{n+1},Y_{n}) \right]
\Pr[\rd \tilde{W}^\mathfrak{a}_{n}=\rd W^{\mathfrak{a}}(Y_{n+1},Y_{n})],
\end{align}
where
\begin{align}
T^{\mathfrak{a}}{}_{i}(Y_{n+1},Y_{n}):=
\pfrac{}{\breve{k}^{i}_{n+1}}\rd W^{\mathfrak{a}}(Y_{n+1},Y_{n}),
\end{align}
and $y_{n+1}$ appearing on the right hand side is regarded as a function 
{$y_{n+1}(k_{n+1},Y_{n})$} determined by eq.~\eqref{def-h}. 
Using the implicit relationship \eqref{def-h} between $y_{n+1}$ and $k_{n+1}$, 
we can get the following relation by use of a differentiation with respect to 
$k_{n+1}^i$,
\begin{align}
\left[\delta^{i}{}_{\ell}-\pfrac{\gamma^{-1}}{\hat{{y}}_{n+1}^{\ell}}\rd t\right]
\pfrac{y_{n+1}^{\ell}}{\breve{k}_{n+1}^{j}}
=\left[\frac{1}{2m}\delta^{i}{}_{j}\gamma^{-1}|_{Y_{\bar n}}
+\frac{k^{i}_{\bar{n}}}{2m}\pfrac{\gamma^{-1}}{\breve{{k}}^{j}_{n+1}} \right]\rd t,
\end{align}
which indicates that $\partial y^{\ell}_{n+1}/\partial\breve{k}^j_{n+1}\sim o(\rd t)$. 
Therefore,
\begin{align}
&T^{\mathfrak{a}}{}_{i}({Y_{n+1},Y_{n}})
=\pfrac{\rd W^{\mathfrak{a}}}{\breve{k}^{i}_{n+1}}
+\pfrac{y^{k}_{n+1}}{\breve{k}^{i}_{n+1}}
\pfrac{\rd W^{\mathfrak{a}}}{\hat{y}^{k}_{n+1}}
=\pfrac{\rd W^{\mathfrak{a}}}{\breve{k}^{i}_{n+1}}+o(\rd t^{2})\notag\\
&\qquad=(\hat{\mathcal{R}}^{-1})^{\mathfrak{a}}{}_{i}|_{Y_{\bar{n}}}
+\frac{1}{2}\left\{m\gamma^{-1}\pfrac{}{\breve{k}^{i}}
(\hat{ \mathcal{R}}^{-1})^{\mathfrak{a}}{}_{j}\mathcal{A}^{j}_{n} 
-\pfrac{}{\breve{k}^{i}}[(\hat{ \mathcal{R}}^{-1})^{\mathfrak{a}}{}_{j} F^{j}]
-\pfrac{}{\breve{k}^{i}}[(\hat{ \mathcal{R}}^{-1})^{\mathfrak{a}}{}_{j}\bar F^{j}]
\right\}_{Y_{\bar{n}}}\rd t,
\end{align}
where terms of order $o(\rd t^{2})$ and higher have been omitted. 
Finally, the transition probability of the forward process can be expressed as
\begin{align}
&\Pr[\tilde{Y}_{n+1}=Y_{n+1}|\tilde{Y}_{n}=Y_{n}]\notag\\
&\qquad=\left[1+\frac{k^{i}}{2m}\gamma^{-2}\pfrac{}{\hat{y}^{i}}\gamma \rd t
+o(\rd t^{2}) \right]_{Y_{\bar{n}}}
\left|\frac{k_{0}\partial_{0} t}{m\lambda g} \right|_{Y_{n+1}}\Delta(Y_{n+1},Y_{n})
\det\left[T^{a}{}_{i}(Y_{n+1},Y_{n}) \right]\notag\\
&\qquad\quad\times \Pr[\rd \tilde{W}^\mathfrak{a}_{n}
=\rd W^{\mathfrak{a}}{(Y_{n+1},Y_{n})}].
\end{align}
The single step transition probability in the reversed process can be 
evaluated following a similar procedure, yielding
\begin{align}
&\Pr[\tilde{Y}^{-}_{N-n}=I(Y_{n})|\tilde{Y}^{-}_{N-n-1}=I(Y_{n+1})]\notag\\
&\qquad=\left[1+\frac{k^{i}}{2m}\gamma^{-2}\pfrac{}{\hat{{y}}^{i}}\gamma \rd t
+o(\rd t^{2}) \right]_{I(Y_{\bar{n}})}
\left|\frac{k_{0}\partial_{0} t}{m\lambda g} \right|_{I(Y_{n})}\notag\\
&\qquad\quad\times \Delta(I(Y_{n}),I(Y_{n+1}))
\det\left[T^{a}{}_{i}(I(Y_{n}),I(Y_{n+1})) \right]
\Pr[\rd \tilde{W}^{\mathfrak{a}}_{n}=\rd W^{\mathfrak{a}}(I(Y_{n}),I(Y_{n+1}))]\notag\\
&\qquad=\left[1-\frac{k^{i}}{2m}\gamma^{-2}\pfrac{}{\hat{y}^{i}}\gamma \rd t
+o(\rd t^{2}) \right]_{Y_{\bar{n}}}
\left|\frac{k_{0}\partial_{0} t}{m\lambda g} \right|_{Y_{n}}\Delta(Y_{n+1},Y_{n})
\det\left[T^{a}{}_{i}(I(Y_{n}),I(Y_{n+1})) \right]\notag\\
&\qquad\quad\times \Pr[\rd \tilde{W}^{\mathfrak{a}}_{n}=\rd W^{\mathfrak{a}}
(I(Y_{n}),I(Y_{n+1}))].
\end{align}
The Jacobian matrix $T^{a}{}_{i}(I(Y_{n}),I(Y_{n+1}))$ in the reversed process reads
\begin{align}
&T^{\mathfrak{a}}{}_{i}(I(Y_{n}),I(Y_{n+1}))\notag\\
&\qquad=(\hat{\mathcal{R}}^{-1})^{\mathfrak{a}}{}_{i}|_{I(Y_{\bar{n}})}
+\frac{1}{2}\left\{m\gamma^{-1}\pfrac{}{\breve{k}^{i}}
(\hat{ \mathcal{R}}^{-1})^{\mathfrak{a}}{}_{j}\mathcal{A}^{j}_{n} 
-\pfrac{}{\breve{k}^{i}}[(\hat{\mathcal{R}}^{-1})^{\mathfrak{a}}{}_{j} F^{j}]
-\pfrac{}{\breve{k}^{i}}[(\hat{\mathcal{R}}^{-1})^{\mathfrak{a}}{}_{j}\bar F^{j}]
\right\}_{I(Y_{\bar{n}})}\rd t \notag\\
&\qquad=(\hat{ \mathcal{R}}^{-1})^{\mathfrak{a}}{}_{i}|_{Y_{\bar{n}}}
+\frac{1}{2}\left\{-m\gamma^{-1}\pfrac{}{\breve{k}^{i}}
(\hat{\mathcal{R}}^{-1})^{\mathfrak{a}}{}_{j}\mathcal{A}^{j}_{n} 
+\pfrac{}{\breve{k}^{i}}[(\hat{\mathcal{R}}^{-1})^{\mathfrak{a}}{}_{j} F^{j}]
-\pfrac{}{\breve{k}^{i}}[(\hat{\mathcal{R}}^{-1})^{\mathfrak{a}}{}_{j}\bar F^{j}]
\right\}_{Y_{\bar{n}}}\rd t.
\end{align}

In the continuum limit, we have
\begin{align}\label{rate-T}
&\lim_{N\rightarrow +\infty}\prod_{n=0}^{N-1}\frac{\det[T^{a}{}_{i}(Y_{n+1},Y_{n+1})]}
{\det[T^{a}{}_{i}(I(Y_{n}),I(Y_{n+1})) ]}\notag\\
&\qquad=\exp\left[\int_{t_{I}}^{t_{F}}\hat{\mathcal{R}}^{i}{}_{\mathfrak{a}}
\left\{m\gamma^{-1}\pfrac{}{\breve{k}^{i}}(\hat{ \mathcal{R}}^{-1})^{\mathfrak{a}}{}{j}
\mathcal{A}^{j}
-\pfrac{}{\breve{k}^{i}}[(\hat{ \mathcal{R}}^{-1})^{\mathfrak{a}}{}_{j}\bar F^{j}] 
\right\} \rd t\right]\notag\\
&\qquad=\exp\left[\int_{t_{I}}^{t_{F}}\gamma^{-1}\hat{\mathcal{R}}^{i}{}_{\mathfrak{a}}
\pfrac{}{\breve{k}^{i}}(\hat{ \mathcal{R}}^{-1})^{\mathfrak{a}}{}_{j}
[m a^{j}-\mathcal{F}^{j}_{\text{em}}]\rd t\right]
\exp\left[-\int_{t_{I}}^{t_{F}}\pfrac{}{\breve{k}^{i}}F^{i} \rd t\right],
\end{align}
where the exponent in the last term can be expanded as
\begin{align}
-\pfrac{}{\breve{k}^{i}}F^{i}=&-\(\mathcal{F}^{i}_{\text{em}}
-\frac{1}{m}\varGamma^{i}{}_{\alpha\beta}k^{\alpha} k^{\beta} \)
\pfrac{}{\breve{k}^{i}}\gamma^{-1}
+\gamma^{-1}\frac{k^{\mu}}{m}g^{-1}\pfrac{}{y^{\mu}}g\notag\\
&-\frac{\gamma^{-1}}{k_{0}}\left[\frac{1}{m}\partial_{0} 
g_{\alpha\beta}k^{\alpha} k^{\beta}+(\mathcal{F}_{\text{em}})_{0} \right].
\end{align}
Meanwhile, we also have
\begin{align}\label{rate-gamma}
&\lim_{N\rightarrow +\infty}\prod_{n=0}^{N-1}
\frac{\left[1+\frac{k^{i}}{2m}\gamma^{-2}\pfrac{}{\hat{y}^{i}}\gamma \rd t
+o(\rd t^{2}) \right]_{Y_{\bar{n}}}}
{\left[1-\frac{k^{i}}{2m}\gamma^{-2}\pfrac{}{\hat{y}^{i}}\gamma \rd t
+o(\rd t^{2}) \right]_{Y_{\bar{n}}}}
=\exp\left[\int_{t_{I}}^{t_{F}}\gamma^{-2}\frac{k^{i}}{m}\pfrac{}{\hat{y}^{i}}
\gamma\rd t \right].
\end{align}
Moreover, the continuum limit of the ratio of the probabilities of the 
Gaussian increments reads
\begin{align}\label{rate-dw}
&\lim_{N\rightarrow +\infty}\prod_{n=0}^{N-1}
\frac{\Pr[\rd \tilde{W}^\mathfrak{a}_{n}=\rd W^\mathfrak{a}(Y_{n+1},Y_{n})]}
{\Pr[\rd \tilde{W}^\mathfrak{a}_{n}=\rd W^\mathfrak{a}(I(Y_{n}),I(Y_{n+1}))]}
=\exp\left[\int_{t_{I}}^{t_{F}}2\gamma^{-2}(\hat{\mathcal{D}}^{-1})_{ij}
(m\mathcal{A}^{i}-\gamma F^{i})\gamma\bar{F}^{j}\rd t \right],
\end{align}
where the integrads can also be expanded as
\begin{align}
&2\gamma^{-2}(\hat{\mathcal{D}}^{-1})_{ij}(m\mathcal{A}^{i}-\gamma F^{i})\gamma\bar F^{j}
\notag\\
&\qquad=2 \gamma^{-1} (\mathcal{D}^{-1})_{ij} (m a^{i} - \mathcal{F}^{i}_{\text{em}})
\left( \mathcal{F}^{j}_{\text{dp}}+\mathcal{F}^{j}_{\text{add}}
-\frac{1}{2}\mathcal{D}^{jk} \gamma^{1/2} \nabla_{k}^{(h)} \gamma^{-1/2} \right)\notag\\
&\qquad=\gamma^{-1} \frac{1}{T_{\rm B}} (m a^{\mu} -\mathcal{F}^{\mu}_{\text{em}}) U_{\mu} 
+ \gamma^{-1} (\mathcal{R}^{-1})^a _j
\nabla_{i}^{(h)} \mathcal{R}^{i}{}_{a} (m a^{j} - \mathcal{F}^{j}_{\text{em}}) \notag\\
&\qquad\quad +\frac{1}{2} \gamma^{-2} (m a^{i} - \mathcal{F}^{i}_{\text{em}}) 
\nabla_{i}^{(h)}\gamma\notag\\
&\qquad=\gamma^{-1} \frac{1}{T_{\rm B}}  (m a^{\mu} - \mathcal{F}^{\mu}_{\text{em}}) U_{\mu} 
+ \gamma^{-1} (\hat{\mathcal{R}}^{-1})^{a}{}_{j}
\pfrac{}{\breve{k}^{i}} \hat{\mathcal{R}}^{i}{}_{a} (m a^{j} - \mathcal{F}^{j}_{\text{em}}) 
\notag\\
&\qquad\quad+\gamma^{-2} (m a^{i} - \mathcal{F}^{i}_{\text{em}})
\pfrac{}{\breve{k}^{i}}\gamma
-\frac{\gamma^{-1}}{k_{0}}(ma_{0}-(\mathcal{F}_{\text{em}})_{0}).
\end{align}

In the end, by combining all the continuum limit together, we obtain the ratio of the 
conditional probabilities for the forward and reversed processes,
\begin{align}
&\lim_{N\rightarrow +\infty}\prod_{n=0}^{N-1}
\frac{\Pr[\tilde{Y}_{n+1}=Y_{n+1}|\tilde{Y}_{n}=Y_{n}]}
{\Pr[\tilde{Y}^{-}_{N-n}=I(Y_{n})|\tilde{Y}^{-}_{N-n-1}=I(Y_{n+1})]}\notag\\
&\qquad=\frac{\lambda|_{Y_{0}}}{\lambda|_{Y_{N}}}\times\text{eq.~}
\eqref{rate-p0}\times \text{eq.~}\eqref{rate-g} \times \text{eq.~} 
\eqref{rate-t0} \times\text{eq.~}\eqref{rate-T} \times\text{eq.~}
\eqref{rate-gamma} \times\text{eq.~}\eqref{rate-dw} \notag\\
&\qquad=\frac{\lambda|_{Y_{0}}}{\lambda|_{Y_{N}}}\exp\left[\int_{t_{I}}^{t_{F}}
\gamma^{-1}\frac{1}{T_{\rm B}} (m a^{\mu} -\mathcal{F}^{\mu}_{\text{em}}) 
U_{\mu} \rd t\right]\notag\\
&\qquad\quad\times\exp\left[\int_{t_{I}}^{t_{F}} 
\gamma^{-2}\(\frac{k^{i}}{m}\pfrac{}{\hat{y}^{i}}\gamma
+\gamma\pfrac{}{t}\gamma+m \mathcal{A}^{i}\pfrac{}{\breve{k}^{i}}\gamma \)\rd t\right]
\notag\\
&\qquad=\frac{\lambda|_{Y_{0}}}{\lambda|_{Y_{N}}}
\exp\left[\int_{t_{I}}^{t_{F}}\gamma^{-1}\frac{1}{T_{\rm B}} 
(m a^{\mu} -\mathcal{F}^{\mu}_{\text{em}}) U_{\mu} \rd t\right]
\exp\left[\int_{t_{I}}^{t_{F}} \gamma^{-2}\mathscr{V}(\gamma)\rd t\right]\notag\\
&\qquad=\frac{(\lambda\gamma^{-1})|_{Y_{0}}}{(\lambda\gamma^{-1})|_{Y_{N}}}
\exp\left[\int_{t_{I}}^{t_{F}}\gamma^{- 1}\frac{1}{T_{\rm B}} 
(m a^{\mu} -\mathcal{F}^{\mu}_{\text{em}}) U_{\mu} \rd t\right],
\end{align}
which is exactly eq.~\eqref{continuum-limit} in the main text.

\end{appendices}

\providecommand{\href}[2]{#2}\begingroup
\providecommand{\eprint}[2][]{\href{http://arxiv.org/abs/#2}{arXiv:#2}}

\end{document}